\documentclass[twocolumn]{revtex4-2}
\usepackage{graphicx}
\usepackage{amsmath,amssymb}
\usepackage{physics,siunitx}
\usepackage{pgfplotstable}
\usepackage{booktabs}
\usepackage{graphicx}
\usepackage{float}
\usepackage[english]{babel}
\usepackage{appendix}

\usepackage[normalem]{ulem}

\begin{document}

\title{Macroscopic coherence and vorticity in room-temperature polariton condensate confined in a self-assembled perovskite microcavity}

\author{Martin Montagnac$^{1,\dagger}$}
\author{Yesenia A. García Jomaso$^{1,\dagger}$}
\author{Emiliano Robledo Ibarra$^1$}
\author{Rodrigo Sánchez-Martínez$^1$}
\author{Moroni Santiago García$^2$}
\author{César L. Ordóñez-Romero$^1$}
\author{Hugo A. Lara-García$^{1,*}$}
\author{Arturo Camacho-Guardian$^{1,*}$}
\author{Giuseppe Pirruccio$^{1,*}$}

\affiliation{1 Instituto de Física, Universidad Nacional Autónoma de México, 
Apartado Postal 20-364, Ciudad de México C.P. 01000, México}
\affiliation{2 Instituto Nacional de Astrof\'isica, \'Optica y Electr\'onica, 
Calle Luis Enrique Erro 1, Sta. Ma. Tonantzintla, Puebla CP 72840, M\'exico}

 \thanks{$^{\dagger}$These authors contributed equally to this work.}
\date{\today}
\maketitle

\section*{Abstract}

Exciton-polariton Bose-Einstein condensation at room temperature offers a promising pathway toward quantum photonic technologies that can operate under ambient conditions. A key challenge in this field is to engineer a controlled platform where strong confinement, nonlinear interactions, and structural disorder coexist, unlocking access to rich collective behavior and unconventional condensate dynamics. We demonstrate polariton condensation in CsPbBr$_3$ microplatelets that self-assemble into whispering gallery mode microresonators featuring tight lateral photon confinement finely balanced with intrinsic disorder. The system exhibits hallmark signatures of out-of-equilibrium condensation, including a non-linear increase in emission intensity, spectral narrowing, and interaction-induced blueshift.
Intrinsic disorder subtly reshapes the cavity energy landscape, inducing condensate fragmentation and enabling direct optical access to the condensate wavefunction. Interferometric measurements reveal extended phase coherence, whereas characteristic fork-shaped fringe dislocations confirm the presence of quantized vortices pinned by the disordered potential. These topological excitations underscore the rich physics driven by the interplay of gain, loss, confinement, and disorder.
Our work establishes a scalable platform for investigating driven-dissipative quantum fluids of light at room temperature, where the intrinsic disorder balances optical confinement and provides a window into condensate wavefunction, coherence, and vortex phenomena. This study system opens new opportunities for exploring many-body physics and potentially advancing topological photonics in integrable microcavity architectures.


\section*{Introduction} Quantum vortices are topological defects that emerge in diverse systems of coherent quantum matter governed by phase rigidity and nonlinear interactions, including superfluids, superconductors, and Bose--Einstein condensates (BECs)~\cite{RevModPhys.38.298,RevModPhys.66.1125,Weiler2008}. These quantized vortices carry discrete angular momentum and arise either spontaneously~\cite{Weiler2008} or deterministically through external perturbations~\cite{Matthews1999}. They manifest as phase singularities in the macroscopic wavefunction, typically forming under rotation or applied magnetic fields. The dynamics and interactions of vortices play a central role in dissipationless transport~\cite{Chien2015}, critical behavior, and the onset of quantum turbulence~\cite{Henn2009,Navon2016,Dogra2023}. In atomic BECs, the controlled generation of vortex lattices in rapidly rotating quantum gases has enabled fundamental studies of quantum fluid dynamics and topological phase transitions~\cite{Cooper2001,cooper2008rapidly}. Such systems provide a versatile platform to engineer and detect strongly correlated topological states, including Laughlin states, through the interplay of rotation, interactions, and quantum coherence~\cite{Abo2001,Mukherjee2022,Leonard2023,Lunt2024,Yao2024}.

Exciton-polaritons, hybrid quasiparticles arising from the strong coupling between excitons and photons, have emerged as a powerful platform to realize quantum fluids~\cite{Deng2010,Carusotto2013,Keeling01032011,Byrnes2014}. Due to their ultrasmall effective mass inherited from the photonic component, they can undergo Bose–Einstein condensation even at room temperature~\cite{guillet2016polariton,Su2020}. This property has opened the door to studying superfluidity~\cite{Lerario2017,Peng2022}, quantum vortices~\cite{LagoudakisHalfQuatumVortices,Lagoudakis2008,Sanvitto2010,Lagoudakis2011,Nardin2011,Sanvitto2016,Tosi2012}, polariton transistors~\cite{Zasedatelev2019}, and strong photon–photon interactions~\cite{Sun2017,Takemura2014}. These features render polaritons a tunable system for exploring unprecedented quantum many-body phases of light and matter, and position them as promising candidates for designing optoelectronic devices with ultrafast response times.

In driven-dissipative systems, such as polariton condensates, vortices arise from the intricate balance of gain, loss, non-linearity, and spatial inhomogeneity~\cite{Lagoudakis2008,Lagoudakis2011}. Unlike equilibrium systems, they can spontaneously form without imposed rotation, often seeded by structural disorder~\cite{Lagoudakis2011}, optical pumping~\cite{Berger2020,Sanvitto2010}, or fluctuations near the condensation threshold. Vortex dynamics and diffusion in these systems have been investigated in connection with turbulent behavior and Kolmogorov-like energy cascades~\cite{Panico2023,Comaron2025}. More recently, turbulence-like phenomena in confined geometries have opened new perspectives, where nonlinearity and spatial boundaries interplay in non-trivial ways. Beyond fundamental interest, vortices have also found applications in photonics, where their topological charge serves as a robust degree of freedom for encoding and manipulating information~\cite{Drori2023,Huang2025}.

Exciton-polariton condensates provide a rich platform to study such non-equilibrium quantum hydrodynamics. However, most experimental realizations have relied on cryogenic temperatures and epitaxially grown Fabry-Pérot microcavities, limiting both accessibility and tunability. This has motivated the search for alternative materials that enable polariton condensation under ambient conditions.

Metal halide perovskites, and in particular CsPbBr$_3$~\cite{Swarnkar2016,Shamsi2019,yu2020perovskite}, have recently emerged as ideal candidates for this purpose due to their large exciton binding energy~\cite{Protesescu2015,Becker2018}, strong photoluminescence~\cite{Yakunin2015,Utzat2019,Raino2022}, and compatibility with low-cost solution processing~\cite{Mendez2025}. Polariton lasing and condensation have already been demonstrated in one-dimensional Fabry–Pérot cavities using CsPbBr$_3$~\cite{Georgakilas2025}. In parallel, whispering gallery mode (WGM) microcavities based on CsPbBr$_3$ have only recently been realized and offer unique advantages: they support polariton modes with non-trivial spatial structure and tight lateral confinement, without the need for lithographic patterning~\cite{Polimeno2024}.

In this article, we demonstrate room-temperature polariton condensation in CsPbBr$_3$ microplatelets that act as self-assembled,  WGM resonators. These cavities feature hard spatial boundaries and strong lateral confinement, creating an environment in which gain, loss, and nonlinear interactions are intricately intertwined, giving rise to dynamical regimes inaccessible in weakly confined or near-equilibrium systems. Strong photon confinement is achieved via high-quality microcrystals that naturally sustain high-quality WGMs, enhancing light-matter coupling and promoting condensation.

The presence of an unavoidable structural disorder, which is often seen as a drawback, can instead be harnessed as a powerful probe of the internal structure of the condensate. It locally perturbs the cavity energy landscape inducing partial fragmentation of the condensate and it enables coupling between the otherwise bound WGM and free-space radiation via light scattering, thereby allowing direct real-space access to the condensate wavefunction and its phase properties.

We first observe the characteristic hallmarks of polariton condensation, including a non-linear emission threshold (polariton lasing), spectral narrowing, and an interaction-induced blueshift. Beyond these standard indicators, we detect the emergence of off-diagonal long-range order (ODLRO), evidenced by interferometric measurements of the first-order spatial coherence function, $g_1(\mathbf{r}, -\mathbf{r})$, which reveals phase coherence extending over several microns. Strikingly, the interferograms exhibit fork-like dislocations in the interference fringes, clear signatures of quantized vortices in the condensate wavefunction appearing as spontaneously formed vortices pairs. These topological defects emerge without any external stirring or phase imprinting, reflecting the intrinsic tendency of the condensate to develop phase singularities due to spatial confinement and disorder.

Our results provide direct evidence of the macroscopic coherence and topological nature of perovskite WGM polariton condensates, unveiling new opportunities for the exploration of room-temperature many-body polariton physics in high-quality, self-assembled cavities.

\begin{figure}[h]
    \centering
    \includegraphics[width=0.5\textwidth]
    {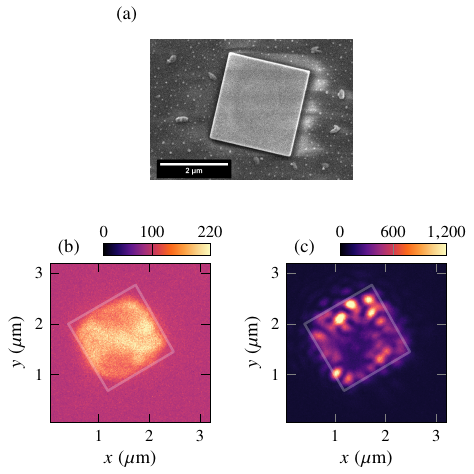}
  \caption{\textbf{Growth, morphology, and photoluminescence of CsPbBr$_3$ microplatelets.} 
(\textbf{a}) Scanning electron microscopy (SEM) image of CsPbBr$_3$ microplatelets exhibiting a well-defined square morphology with sharp edges and smooth facets. Step-like terraces reveal anisotropic growth along specific crystallographic directions. 
Real-space PL intensity profile of a microplatelet for incident power below (\textbf{b}) and above (\textbf{c}) the condensation threshold.}
    \label{fig:growth}
\end{figure}

\begin{figure*}[htbp]
    \centering
    \includegraphics
    {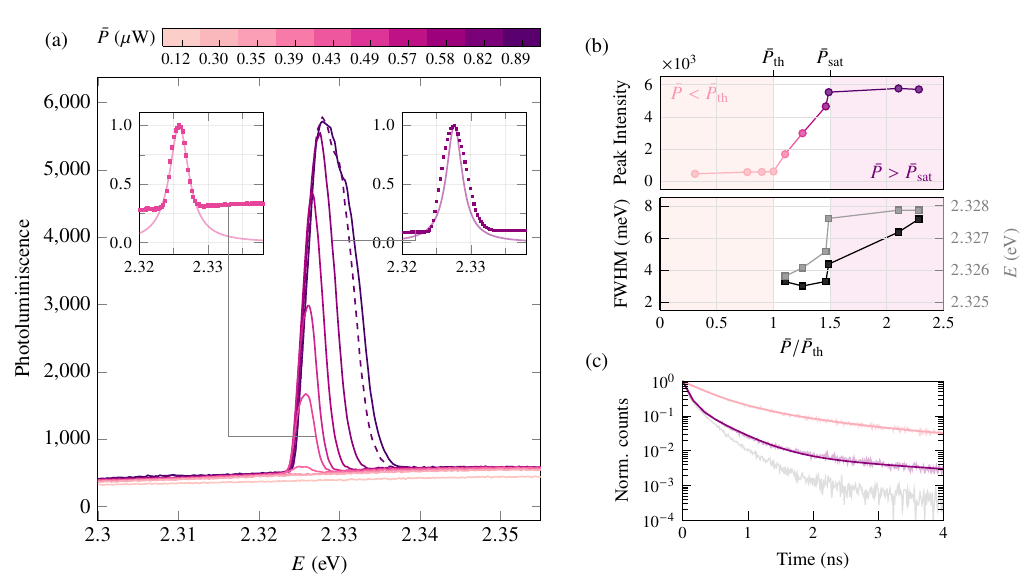}
    \caption{\textbf{Polariton condensation threshold and nonlinear emission.} 
    (a) Power-dependent PL spectra show a sudden increase in intensity near $P_{\text{th}} \approx 0.40\,\mu\text{W}$, signaling the onset of polariton condensation. Insets show the Lorentzian fit of the spectral lineshape for two values of the pump power.
    (b) Upper panel: Peak intensity of the polariton laser spectra as a function of excitation power, exhibiting a nonlinear increase followed by saturation around $P_{\text{sat}} \approx 0.60\,\mu\text{W}$. 
    (b) Lower panel: Full width at half maximum (FWHM) and peak emission energy vs. pump power.
    (c) \textbf{Time-resolved photoluminescence across condensation threshold.} Lifetime measurement below (pale pink) and above (purple) condensation threshold.
    The instrument response function is shown in gray.
    }
    \label{fig:threshold}
\end{figure*}

\section*{ Growth of self-assembled CsPbBr$_3$ WGM microcavities.}
High-quality CsPbBr$_3$ microplatelets were synthesized using a custom-built chemical vapor deposition (CVD) method that yields well-defined two-dimensional crystals with lateral sizes typically between 2 and 4~µm.
The resulting crystals exhibit uniform square-like geometries, flat top surfaces, and intense green photoluminescence under UV illumination, indicating high optical quality and high crystallinity.

Figure~\ref{fig:growth}(a) displays the scanning electron microscopy (SEM) image of a representative microcrystal, revealing its well-defined square geometry, sharp edges, and smooth facets.
The presence of step-like terraces indicates a layer-by-layer growth mechanism.
The corresponding two-dimensional atomic force microscopy (AFM) topographic map of a representative microplatelet can be found in Figure~S1, where the corresponding height profile confirms a uniform thickness of approximately $200\,\text{nm}$ and a nanometrically smooth surface.
The alignment of the terraces with the underlying crystallographic axes reflects the anisotropic growth characteristic of high-quality single crystals.
The combination of high refractive index (Figure~S3), planar geometry, and low surface roughness makes these microplatelets an ideal platform for the formation of self-assembled, high-$Q$ optical cavities capable of supporting WGMs.

Importantly, the optical response of these microplatelets is governed by strong excitonic effects at room temperature. Absorption spectra reveal a steep onset near $510~\text{nm}$, accompanied by a pronounced shoulder attributed to excitonic resonances and indicative of the high crystalline quality of the material. Photoluminescence spectra show narrowband emission centered around $530~\text{nm}$ with a Stokes shift of approximately $10\,\text{nm}$.
The presence of such a robust excitonic feature combined with high-Q WGMs, provides the necessary conditions for achieving strong light--matter coupling and polariton condensation under ambient conditions.

\section*{Polariton lasing}

The photoluminescence (PL) spectrum of a single microcrystal is measured with a diffraction-limited optical microscope under pulsed illumination. The pump pulses have a temporal duration of approximately 100\,fs and a repetition rate of $50\,\text{KHz}$.
Non-resonant excitation is performed using a $\lambda_{p} = 460\,\text{nm}$ and adjusting the diameter of the excitation spot to the size of the crystal.

We begin by showing the microscope images of the real-space PL under different pumping conditions, and then we analyze the spectral properties of the emitted light.
Figure~\ref{fig:growth}(c) shows the PL image obtained for a pump fluence well below the lasing threshold.
Incoherent light is emitted throughout the whole crystal with a slightly non-uniform emission intensity signaling an inhomogeneous energy landscape, this likely  due to a small amount of disorder sensed by polaritons. 
In Figure~\ref{fig:growth}(d) we show the emission from the same crystal obtained for a fluence above the lasing threshold. In contrast to the incoherent emission, the lasing mode profile is strongly spatially structured and asymmetric, characterized by isolated regions of high polariton density. We attribute this fragmentation to the interplay between nonlinearities and disorder. The presence of a small degree of disorder within the cavity is confirmed by the non-uniform spatial emission seen below threshold \cite{Lagoudakis2008,Lagoudakis2011}.
Disorder modifies the energy landscape felt by polaritons that condense in multiple local minima \cite{Lagoudakis2008,Lagoudakis2011}. 

In Figure~\ref{fig:threshold}(a), we plot the emission spectrum as a function of the average pump power measured at the sample position.
A clear lasing peak appears on top of the incoherent linear PL spectrum for values above a threshold power $\bar{P}_\mathrm{th} \approx 0.40\,\mu\text{W}$.
This peak undergoes the typical blueshift found in polariton condensates and is accompanied by a spectral narrowing down to a full width at half maximum (FWHM) value of approximately $3\,\text{meV}$.
Both quantities are plotted as a function of the average pump power in Figure~\ref{fig:threshold}(b). 

For intermediate average powers, $\bar{P}\sim 0.40-0.60\,\mu\text{W}$, the blueshift of the laser peak stays linear and equal to $\Delta E\approx\,3\text{meV}$, while the FWHM remains essentially constant. In this region, the polariton laser spectrum gradually deviates from a Lorentzian shape, as shown in the insets of Figure~\ref{fig:threshold}(a) where we compare the normalised laser spectrum with a Lorentzian function centered around the polariton energy $E_{\text{P}}$ and having a fixed broadening of $\gamma_{\text{P}}$. 
The asymmetric broadening of the laser peak resulting in a high-energy long tail signals polariton interactions~\cite{porras} and it is one of the characteristics of polariton condensates. As power further increases above $\bar{P}\sim 0.60\,\mu\text{W}$, intensity and polariton energy saturation occur, together with a sudden increase of the FWHM, as can be seen in Figure~\ref{fig:threshold}(b), upper panel.
The appearance of a jump in all these three quantities corresponds to the emergence of an additional peak in the PL spectrum, which is clearly distinguished in the dashed curve in Figure~\ref{fig:threshold}(a). 
By performing measurements on many other similar microcrystals, we observed that this peak always appears on the high-energy side of the main peak. We speculate that this extra peak may be of the same WGM order associated to the polariton laser but whose resonant wavelength is slightly modified by the presence of disorder and therefore requires a larger energy and fluence to be populated. 

\begin{figure}[ht]
    \centering
    \includegraphics[width=\linewidth]
    {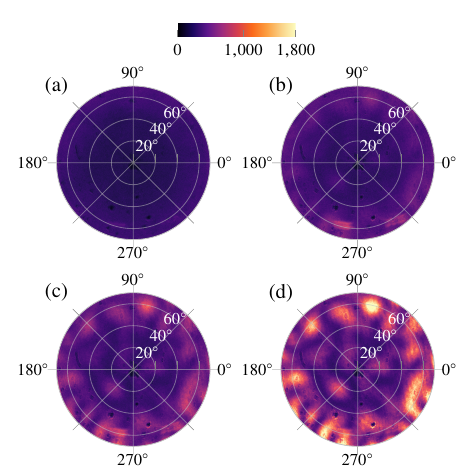}
    \caption{\textbf{Momentum-space evolution of polariton emission across the condensation threshold.} Angular-resolved PL maps for increasing powers. Below threshold (a-b), the emission intensity is isotropic and low. As the pump power goes above the condensation threshold (c-d), several intensity lobes appear at large elevation angles associated with radiative modes outcoupled from the cavity with well defined wavevectors.}
    \label{fig:angular_modes}
\end{figure}

Another signature of polariton lasing is the accelerated PL decay above the condensation threshold.
The time-resolved photoluminescence (see Methods) is plotted in Figure \ref{fig:threshold}~(c) for different pump fluence.
Below threshold (pink points), the decay is bi-exponential with characteristic times $\tau_1^{\text{below}} = 363\,\text{ps}$ and $\tau_2^{\text{below}} = 1\,\text{ns}$ with corresponding amplitudes $0.65$, and $0.28$.
Above threshold (purple points), PL decays much faster, with $\tau_1^{\text{above}} = 86\,\text{ps}$ and amplitude $0.75$, reflecting a rapid bosonic stimulated scattering into the ground state of the condensate followed by fast leakage through the cavity governed by the short cavity photon lifetime.
The second exponential contributes with an approximately unchanged $\tau_2^{\text{above}} = 405\,\text{ps}\sim \tau_1^{\text{below}}$ and amplitude of $0.24$.
These decay times are in the order of those already reported in the literature\cite{Polimeno2024,Becker2018}.
Even though the measured condensate lifetime is limited by the instrument response function (IRF) of our time-correlated single-photon set-up (shown in gray in the Figure), the observed lifetime shortening corroborates the formation of a condensed phase and highlights the potential very fast response of perovskite polariton condensate systems.
The second decay constant remains similar below and above the threshold, suggesting that it is associated with remaining noncondensed polaritons.

\section*{Modal decomposition in momentum-space} To gain insight into the modal structure of the PL emission, we perform angle-resolved PL imaging (see Methods). This allows us to obtain a momentum-space picture of the emission by imaging the radiative modes associated with the polariton laser. In Figure~\ref{fig:angular_modes}(a)-(b) the emission appears isotropic and with an almost constant intensity. The lack of directionality is consistent with the incoherent character of the PL below the condensation threshold. Around $P_{\text{th}},$ shown in Figure~\ref{fig:angular_modes}(b-c) discrete angular lobes emerge at large angles consistent with the mainly in-plane light circulation of the WGM and the coherent emission from the microcrystal. The saturation of the intensity of the radiative modes agrees with the saturation of the intensity seen in the spectrum (see Figure~\ref{fig:threshold}(b)), while the background intensity accounts for the remaining incoherent PL (see Figure~\ref{fig:threshold}(a)) associated to the fraction of non-condensed polaritons. Despite spectral broadening and the emergence of an extra peak in the spectrum, the angular distribution of the light intensity does not change significantly in the saturation region, as seen in Figure~\ref{fig:angular_modes}(d).

Importantly, the structure of the momentum-space images for different powers does not display any specific symmetry. This confirms the presence of disorder in the system and agrees with the fragmented polariton density shown in Figure~\ref{fig:growth}(d). Nonetheless, the presence of well-defined high-intensity regions arranged in a ring-like pattern is a preliminary signature of the long-range spatial coherence present in the condensate comprising phase-locked fragments.

 \begin{figure*}[htbp]
    \centering
    \includegraphics
    {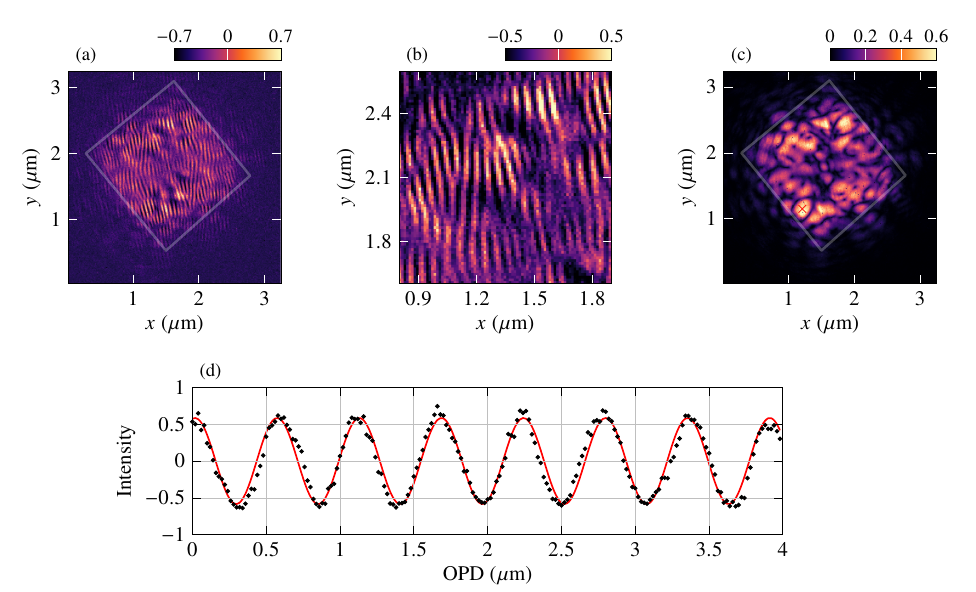}
    \caption{\textbf{Spatial coherence of the polariton condensate.} (a) Interferogram for OPD = $2\ \mu$m showing high-contrast fringes across the whole cavity. Grey box corresponds to the microcrystal area. (b) Close-up view of the interferogram in shown in (a) highlighting the presence of multiple fork-like dislocations.
    (c) Reconstructed spatial map $g^{(1)}(\mathbf{r},-\mathbf{r})$ indicating large extended coherence of the condensate. 
    (d) Fringe visibility for the coordinate marked by a red cross in (c).
    Black dots and red curve are the experimental values and least squares fit, respectively.
    }
    \label{fig:g1}
\end{figure*}

\section*{Macroscopic coherence and off-diagonal long-range order} One of the hallmarks of polariton condensation is its macroscopic spatial coherence extending across the system. The experimental signature of the ODLRO present in the condensate is typically obtained by means of interferometric measurements.
We employ a modified Michelson interferometer where the condensate PL image is overlapped with its retroreflected centrosymmetric copy (see Methods). The schematics of the set-up is shown in Figure S4 of the Supplemental Information.
First, the optimal temporal overlap between the polariton laser pulses is ensured by equaling the optical paths of the two arms of the interferometer.
For a given position of the retroreflector, an interferogram is obtained in which well-defined, high contrast interference fringes are observed across the whole microcrystal, as displayed in Figure~\ref{fig:g1}(a) for an optical path difference OPD = $2\,\mu$m and after normalizing for the spatial-dependent intensity present in the two arms.
This interferogram, result of the integration of approximately 5000  images, i.e., polariton laser pulses, shows that a fixed phase relation exists between diametrically opposite parts of the condensate.
Outside the cavity, marked by a white box, low-intensity interference fringes correspond to the polariton laser light leaking out through the borders of the cavity via surface scattering and creating standing waves upon reflection off other nearby microcrystals facets. We highlight that the overall relative orientation of the fringes seen in Figure~\ref{fig:g1}(a) is not constant throughout the cavity.

Then, scanning the piezomirror over a few micrometers provides the phase advance of the macroscopic condensate wavefunction needed to extract the first-order spatial correlation function, $g^{(1)}(\mathbf r,-\mathbf r)$.
By evolving the condensate phase, the variation of light intensity as a function of the OPD is recorded for each point on the condensate area.
An example of such a curve is shown in Figure~\ref{fig:g1}(d) with black dots and for the coordinate $(x,y) = (1.2,1.2)$, marked with a red cross in Figure~\ref{fig:g1}(c). 
The solid red curve corresponds to the least squares fit function $I(\mathrm{OPD}) = A \cos\left( \frac{2\pi \, \mathrm{OPD}}{\lambda} + \phi \right),
$ where $A$ is the amplitude, $\lambda$ is the central emission wavelength and $\phi$ is the phase offset.
Before fitting, a fast Fourier transform high-pass filter is applied to remove any possible residual offset from the recorded intensity oscillation of each pixel.
The high quality of the fit across the entire area is quantified in the Supplemental Information. The amplitude $A$ corresponds to the first-order correlation function, $g^{(1)}(\mathbf r,-\mathbf r)$ and is plotted in Figure~\ref{fig:g1}(c) for every point in the imaged area.
The autocorrelation point can be observed in the center of the cavity, which is marked by the white box.
The spatial coherence map is structured in domains, some of them clearly encircled by dark closed curves, confirming the partial fragmentation of the condensate.
The large overall values reached by $g^{(1)}(\mathbf r,-\mathbf r)$ imply that disorder-induced modification of the energy landscape is not enough to prevent macroscopic coherence build-up and demonstrates that the ODLRO spatially extends for several micrometers throughout the cavity.
Interestingly, here disorder may also serve as a diagnostic tool by coupling the otherwise confined whispering-gallery modes to propagating modes through light scattering. This coupling enables access to the internal spatial structure of the condensate while preserving its coherence.

\begin{figure*}[t]
    \centering
    \includegraphics
    {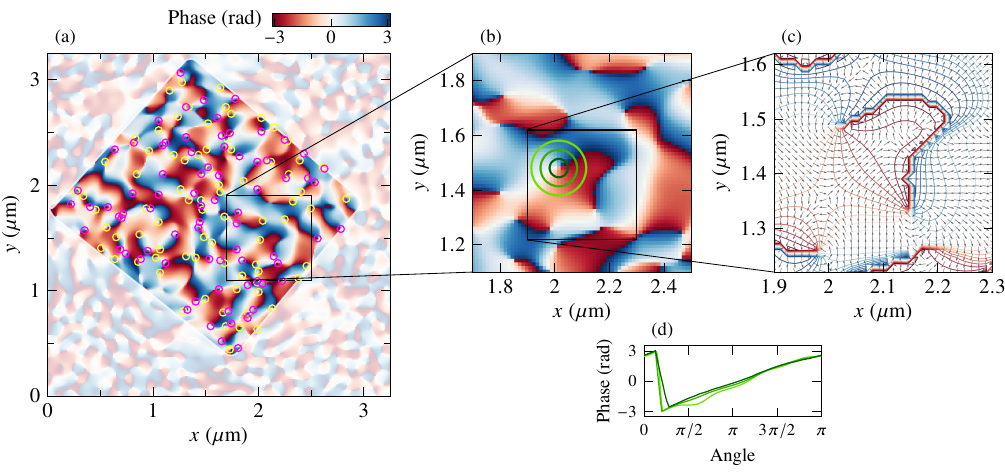}
    \caption{\textbf{Polariton condensate vorticity.}
    (\textbf{a}) Reconstructed phase map of the condensate showing the presence of multiple vortices with yellow and magenta circles.
    (b) Close-up view of (a).
    (\textbf{c}) Velocity field derived from the phase gradient, showing a rotational flow pattern centered on the singularity. The color scale encodes the local phase.
    (\textbf{d}) Phase as a function of angular coordinate along the circular contours of increasing radius overlaid in (b), showing the $2\pi$ winding around the defect core.
    }
    \label{fig:vortices}
\end{figure*}

\section*{Topological defects and quantized vortices}
By taking a closer look at the interferogram shown in Figure~\ref{fig:g1}(a), a non-trivial fringe pattern is evident. Figure~\ref{fig:g1}(b) is a close-up view of Figure~\ref{fig:g1}(a) and shows the presence of multiple fork-like dislocations. In addition, the $g^{(1)}(\mathbf r,-\mathbf r)$ map exhibits several small dark spots, representing defects within areas of otherwise large coherence. 
There, the emitted intensity approaches zero, the phase becomes undefined, and the spatial coherence abruptly drops.
These are signatures of phase singularities, which find their origin in the interplay between polariton interactions and disorder. The phase of the macroscopic condensate wavefunction across the cavity can be retrieved by the off-angle digital holography method (see Methods).
This is shown in Figure~\ref{fig:vortices}(a) where the magenta and yellow circles mark algorithmically searched points where the phase winds and accumulates exactly $\pm 2\pi$, thus corresponding to quantized right- and left-handed quantized vortices with unitary charge. We observe that these features self-organize in vortex-antivortex pairs having opposite handedness, thus maintaining the total angular momentum of the system equal to zero~\cite{Panico2023}.
The observed vortices may form spontaneously, diffuse and eventually get pinned by structural disorder on a time scale faster than the acquisition time of the camera, or they may be associated with the nodes of stationary waves in which polaritons condense. In either case, the temporal integration of the images used to increase the signal-to-noise ratio confirms their stationary character and deterministic evolution. 

To quantitatively confirm the topological charge of these vortices, a close-up view of one of them is shown in Figure~\ref{fig:vortices}(b). We extract the phase as a function of the polar angle along three concentric contours of increasing radius centered around the vortex core. Figure~\ref{fig:vortices}(d) shows the phase wrap of $2\pi$ in all three cases, an almost ideal linear increase, and demonstrates that the quantized circulation is robust against the radial distance from the vortex core. This confirms that the observed defect corresponds to a singly charged vortex, as expected for a fundamental topological excitation in a two-dimensional condensate.

Figure~\ref{fig:vortices}(c) illustrates the gradient of the reconstructed phase field, which is proportional to the velocity field of the polariton condensate. The vector field clearly reveals handed circular flows around vortex cores, consistent with quantized circulation in superfluids, with opposite signs for the vortex–antivortex pairs observed here. This maintains the total average circulation of the field close to zero.

\section*{Conclusion}

In summary, we report the realization of room-temperature polariton condensation in a single perovskite microplatelet acting as an optical cavity. This observation is supported by a comprehensive set of experimental signatures. The emergence of a non-linear emission threshold, spectral narrowing, blueshift, and power saturation all point to the spontaneous formation of a macroscopically occupied polariton state. Spatially resolved interferometry confirms the establishment of long-range phase coherence across the condensate. Moreover, the observation of fork-like dislocations in the interference fringes provides direct evidence for the presence of topological phase singularities, i.e., vortices, of the condensate wavefunction. These phase defects underscore the quantum nature of the macroscopic state and suggest rich dynamical behaviors inherent to two-dimensional quantum fluids. Finally, the abrupt shortening of the emission lifetime above threshold reinforces the interpretation of stimulated scattering into the ground state.

These detailed observations are made possible by the delicate balance between inherent material disorder and high-quality factor optical modes that confine the polariton condensate. In fact, thanks to far-field light scattering, disorder becomes a gateway to deeper understanding, enabling direct imaging of condensate structure, fragmentation, and vortex formation in a scalable material system.

Together, these findings demonstrate a robust platform for investigating quantum collective behavior in light--matter systems under ambient conditions.
This work paves the way toward the development of low-power polaritonic devices and lays the foundation for future explorations of nonequilibrium superfluidity, topological excitations, and quantum turbulence in room-temperature quantum fluids of light.

\section*{Methods-Fabrication-Characterization}

\subsection*{Synthesis of CsPbBr$_3$ microplatelets}
High-quality CsPbBr$_3$ microplatelets were synthesised using a custom-built chemical vapour deposition (CVD) system. A 1:1 molar mixture of CsBr and PbBr$_2$ (total mass 2 mg) was used as the precursor. The powders were placed at the centre of a horizontal tubular furnace and heated to $650\,^\circ\text{C}$. Glass slides coated with a thin SiO$_2$ layer, deposited by dip-coating, were positioned approximately 13 cm downstream from the precursor source, where the local temperature was maintained between 380 and 400 $^\circ$C. The reactor was purged continuously with argon at a flow rate of 30 sccm and kept at a reduced pressure of $\sim0.02\,\text{atm}$ throughout the growth. 
Despite the inherent roughness and low crystallinity of the dip-coated SiO$_2$ substrates, the growth process yielded well-defined microplatelets of CsPbBr$_3$, typically exhibiting square shapes with lateral sizes in the $2-4\,\mu\text{m}$ range.

\subsection {Morphological characterisation}
The morphology and microstructure of the microplatelets were examined by scanning electron microscopy (SEM) using a JEOL JSM-7800F field-emission microscope. High-resolution transmission electron microscopy (HRTEM) was carried out on a JEOL 2010F microscope. Atomic force microscopy (AFM) measurements were performed with an Anton Paar instrument to probe the topography and thickness of individual microplatelets.

\section*{Methods – Optical measurements}
All optical measurements discussed in the main text were performed on a single crystal in a custom-made modular confocal microscope. The schematics of the set-up is shown in Figure. S4 of the Supplemental Information.

The sample is illuminated with laser pulses of duration equal to 100 fs, centered around 460 nm, emitted by an optical parametric amplifier (Orpheus, Light Conversion operated at a repetition rate of 50 KHz) which is pumped by a Yb:KGW laser (Pharos, Light Conversion). The FWHM of the focal point is adjusted to the crystal size and equals approximately 2 µm. 

For the spectral measurements, the collected emitted light is sent to a spectrograph (Kymera 328i, Andor), dispersed by a high-resolution diffraction grating (1200 lines/mm, blazed at 500 nm) and directed to a CCD camera (iDus, Andor). 

The sample is mounted on an xyz linear stage (Microblock, Thorlabs), which provides sufficient alignment precision to position the pump spot at will on the sample. Being the spectral measurement particularly sensitive to the illumination conditions, stability is ensured by checking for mechanical drifts of the sample and measuring the polariton laser spectrum at the beginning and at the end of each experiment.

Angle-resolved emission is performed by imaging the back-focal plane of the high numerical aperture microscope objective (Plan Fluor 100x/0.9 NA, Nikon) onto the entrance slit of the spectrograph coupled to a sCMOS camera (Zyla 4.2P, Andor).

Interferometric measurements are performed by sending the polariton emission collimated by the microscope objective to a Michelson interferometer equipped with a hollow retroreflector (Newport) mounted on a motorized linear stage (V508.951020, Physik Instrumente) and a mirror mounted on a piezostage (Tritor, PiezoJena). The overlapping quasi-collinear beams coming from the two arms follow the same optical path as the one used for the spectral and angular measurements and are imaged on the spectrometer camera. The spatial frequency and average orientation of the fringe pattern result from the relative angle between the two focused beams set by the focal length of tube lens. The temporal overlap on the camera of the two copies of the pulses is obtained by looking for the appearance of interference fringes in the spectrum of the pump laser (Fourier-transform spectroscopy). The computer-controlled phase scan is realized by sinchronizing a 10 nm-step of the piezomirror with the image acquisition and guarantees fast and reliable operation. The interferometer module is enclosed in a box that prevents OPD fluctuations between the two arms during the measurements. 

Lifetime measurements are performed by a home-made time-correlated single-photon counting module coupled to the microscope. We pump the crystal with the same laser used for PL measurements. The pump beam is filtered by a 500 nm long pass filter. The emitted light is directed to a single-photon avalanche photodiode (MPD). The trigger obtained from the Pharos and the signal from the detector are sent to a time-to-digital converter (Time Tagger 20, Swabian Instruments).
All histograms are built with $100\,\text{ps}$ binwidth.
The instrument response function (IRF) has been measured in several ways to verify the consistency of the result and ensure a reliable exponential fit for the short-time dynamics.

\section*{Methods – Vortex analysis}
To retrieve the phase information from the interferogram shown in Figure\ref{fig:vortices} we use the off-angle digital holography method. The two-dimensional fast Fourier transform (FFT) of the interferogram image displays three non-overlapping regions of large intensity. The first one, located at the origin ($k_x=k_y=0$), is associated with the background component, while the other two are located around the finite wavevector ($k_x,k_y$) set by the angle between the interfering beams. Being the interferogram intensity a real quantity, the phase information contained in these two lobes is equivalent and redundant. Therefore, first we use a circular band pass filter centered around one of these two regions and then we perform a carrier shift towards the origin to isolate the phase information from the fringe period. Figure~\ref{fig:vortices}(a) is obtained by inverse FFT of the carrier-shifted image. The filter radius is chosen on the basis of the logarithmic scale of the FFT. Several tests have been run to ensure that the final vortex map is robust against variations of the bandpass filter radius. Consistency has also been checked by comparing the result with elliptical and half-plane-type filters.

\section*{Author contributions}
H.L.-G., A.C.-G., and G.P. conceived and supervised the project. H.L.-G. and E.R.I. designed and fabricated the materials. 
Y.A.G.J., M.M., R.S.-M., C.L.O.-R., and G.P. performed the experiments, analyzed and interpreted the data. A.C.-G., M.S.G., and G.P. carried out the theoretical analysis. 
A.C.-G. and G.P. co-wrote the manuscript with input and comments from all other authors.
Y.A.G.J., M.M. contributed equally.

\section*{Corresponding authors}
Correspondence to H.L. (\href{mailto:hugo.lara@fisica.unam.mx}{hugo.lara@fisica.unam.mx}), 
A.C.G. (\href{mailto:acamacho@fisica.unam.mx}{acamacho@fisica.unam.mx}), 
or G.P. (\href{mailto:pirruccio@fisica.unam.mx}{pirruccio@fisica.unam.mx}).

\section*{ Acknowledgments} We thank Iacopo Carusotto and Pavlos Lagoudakis  for valuable discussions.  The authors acknowledge financial support from the CONAHCYT-Mexico Synergy project 1564464. G.P. acknowledges Grant UNAM DGAPA PAPIIT under Nos. IN104522 and IN104325, CONAHCyT Project Nos. 1564464 and 1098652 and project PIIF 24. H.A.L-G acknowledges financial support from UNAM DGAPA-PAPIIT Grant No. IN106725 and IA107023. The work of Y.A.G J. was supported by the UNAM Posdoctoral Program (POSDOC). R.S.M acknowledges support from CONAHCyT PhD scolarship. C.L.O.R acknowledges financial support from UNAM DGAPA PAPIIT Grant No. IG101424. A.C-G. acknowledges financial support from UNAM DGAPA PAPIIT Grant No. IA101923 and IA101325,  UNAM DGAPA PAPIME Grants No. PE100924 and No. PIIF23 and Project CONAHCYT No. CBF2023-2024-1765.

\bibliography{references}

\begin{thebibliography}{50}%
\makeatletter
\providecommand \@ifxundefined [1]{%
 \@ifx{#1\undefined}
}%
\providecommand \@ifnum [1]{%
 \ifnum #1\expandafter \@firstoftwo
 \else \expandafter \@secondoftwo
 \fi
}%
\providecommand \@ifx [1]{%
 \ifx #1\expandafter \@firstoftwo
 \else \expandafter \@secondoftwo
 \fi
}%
\providecommand \natexlab [1]{#1}%
\providecommand \enquote  [1]{``#1''}%
\providecommand \bibnamefont  [1]{#1}%
\providecommand \bibfnamefont [1]{#1}%
\providecommand \citenamefont [1]{#1}%
\providecommand \href@noop [0]{\@secondoftwo}%
\providecommand \href [0]{\begingroup \@sanitize@url \@href}%
\providecommand \@href[1]{\@@startlink{#1}\@@href}%
\providecommand \@@href[1]{\endgroup#1\@@endlink}%
\providecommand \@sanitize@url [0]{\catcode `\\12\catcode `\$12\catcode
  `\&12\catcode `\#12\catcode `\^12\catcode `\_12\catcode `\%12\relax}%
\providecommand \@@startlink[1]{}%
\providecommand \@@endlink[0]{}%
\providecommand \url  [0]{\begingroup\@sanitize@url \@url }%
\providecommand \@url [1]{\endgroup\@href {#1}{\urlprefix }}%
\providecommand \urlprefix  [0]{URL }%
\providecommand \Eprint [0]{\href }%
\providecommand \doibase [0]{https://doi.org/}%
\providecommand \selectlanguage [0]{\@gobble}%
\providecommand \bibinfo  [0]{\@secondoftwo}%
\providecommand \bibfield  [0]{\@secondoftwo}%
\providecommand \translation [1]{[#1]}%
\providecommand \BibitemOpen [0]{}%
\providecommand \bibitemStop [0]{}%
\providecommand \bibitemNoStop [0]{.\EOS\space}%
\providecommand \EOS [0]{\spacefactor3000\relax}%
\providecommand \BibitemShut  [1]{\csname bibitem#1\endcsname}%
\let\auto@bib@innerbib\@empty
\bibitem [{\citenamefont {Anderson}(1966)}]{RevModPhys.38.298}%
  \BibitemOpen
  \bibfield  {author} {\bibinfo {author} {\bibfnamefont {P.~W.}\ \bibnamefont
  {Anderson}},\ }\bibfield  {title} {\bibinfo {title} {Considerations on the
  flow of superfluid helium},\ }\href
  {https://doi.org/10.1103/RevModPhys.38.298} {\bibfield  {journal} {\bibinfo
  {journal} {Rev. Mod. Phys.}\ }\textbf {\bibinfo {volume} {38}},\ \bibinfo
  {pages} {298} (\bibinfo {year} {1966})}\BibitemShut {NoStop}%
\bibitem [{\citenamefont {Blatter}\ \emph {et~al.}(1994)\citenamefont
  {Blatter}, \citenamefont {Feigel'man}, \citenamefont {Geshkenbein},
  \citenamefont {Larkin},\ and\ \citenamefont {Vinokur}}]{RevModPhys.66.1125}%
  \BibitemOpen
  \bibfield  {author} {\bibinfo {author} {\bibfnamefont {G.}~\bibnamefont
  {Blatter}}, \bibinfo {author} {\bibfnamefont {M.~V.}\ \bibnamefont
  {Feigel'man}}, \bibinfo {author} {\bibfnamefont {V.~B.}\ \bibnamefont
  {Geshkenbein}}, \bibinfo {author} {\bibfnamefont {A.~I.}\ \bibnamefont
  {Larkin}},\ and\ \bibinfo {author} {\bibfnamefont {V.~M.}\ \bibnamefont
  {Vinokur}},\ }\bibfield  {title} {\bibinfo {title} {Vortices in
  high-temperature superconductors},\ }\href
  {https://doi.org/10.1103/RevModPhys.66.1125} {\bibfield  {journal} {\bibinfo
  {journal} {Rev. Mod. Phys.}\ }\textbf {\bibinfo {volume} {66}},\ \bibinfo
  {pages} {1125} (\bibinfo {year} {1994})}\BibitemShut {NoStop}%
\bibitem [{\citenamefont {Weiler}\ \emph {et~al.}(2008)\citenamefont {Weiler},
  \citenamefont {Neely}, \citenamefont {Scherer}, \citenamefont {Bradley},
  \citenamefont {Davis},\ and\ \citenamefont {Anderson}}]{Weiler2008}%
  \BibitemOpen
  \bibfield  {author} {\bibinfo {author} {\bibfnamefont {C.~N.}\ \bibnamefont
  {Weiler}}, \bibinfo {author} {\bibfnamefont {T.~W.}\ \bibnamefont {Neely}},
  \bibinfo {author} {\bibfnamefont {D.~R.}\ \bibnamefont {Scherer}}, \bibinfo
  {author} {\bibfnamefont {A.~S.}\ \bibnamefont {Bradley}}, \bibinfo {author}
  {\bibfnamefont {M.~J.}\ \bibnamefont {Davis}},\ and\ \bibinfo {author}
  {\bibfnamefont {B.~P.}\ \bibnamefont {Anderson}},\ }\bibfield  {title}
  {\bibinfo {title} {Spontaneous vortices in the formation of bose--einstein
  condensates},\ }\href {https://doi.org/10.1038/nature07334} {\bibfield
  {journal} {\bibinfo  {journal} {Nature}\ }\textbf {\bibinfo {volume} {455}},\
  \bibinfo {pages} {948} (\bibinfo {year} {2008})}\BibitemShut {NoStop}%
\bibitem [{\citenamefont {Matthews}\ \emph {et~al.}(1999)\citenamefont
  {Matthews}, \citenamefont {Anderson}, \citenamefont {Haljan}, \citenamefont
  {Hall}, \citenamefont {Wieman},\ and\ \citenamefont
  {Cornell}}]{Matthews1999}%
  \BibitemOpen
  \bibfield  {author} {\bibinfo {author} {\bibfnamefont {M.~R.}\ \bibnamefont
  {Matthews}}, \bibinfo {author} {\bibfnamefont {B.~P.}\ \bibnamefont
  {Anderson}}, \bibinfo {author} {\bibfnamefont {P.~C.}\ \bibnamefont
  {Haljan}}, \bibinfo {author} {\bibfnamefont {D.~S.}\ \bibnamefont {Hall}},
  \bibinfo {author} {\bibfnamefont {C.~E.}\ \bibnamefont {Wieman}},\ and\
  \bibinfo {author} {\bibfnamefont {E.~A.}\ \bibnamefont {Cornell}},\
  }\bibfield  {title} {\bibinfo {title} {Vortices in a bose-einstein
  condensate},\ }\href {https://doi.org/10.1103/PhysRevLett.83.2498} {\bibfield
   {journal} {\bibinfo  {journal} {Phys. Rev. Lett.}\ }\textbf {\bibinfo
  {volume} {83}},\ \bibinfo {pages} {2498} (\bibinfo {year}
  {1999})}\BibitemShut {NoStop}%
\bibitem [{\citenamefont {Chien}\ \emph {et~al.}(2015)\citenamefont {Chien},
  \citenamefont {Peotta},\ and\ \citenamefont {Di~Ventra}}]{Chien2015}%
  \BibitemOpen
  \bibfield  {author} {\bibinfo {author} {\bibfnamefont {C.-C.}\ \bibnamefont
  {Chien}}, \bibinfo {author} {\bibfnamefont {S.}~\bibnamefont {Peotta}},\ and\
  \bibinfo {author} {\bibfnamefont {M.}~\bibnamefont {Di~Ventra}},\ }\bibfield
  {title} {\bibinfo {title} {Quantum transport in ultracold atoms},\ }\href
  {https://doi.org/10.1038/nphys3531} {\bibfield  {journal} {\bibinfo
  {journal} {Nature Physics}\ }\textbf {\bibinfo {volume} {11}},\ \bibinfo
  {pages} {998} (\bibinfo {year} {2015})}\BibitemShut {NoStop}%
\bibitem [{\citenamefont {Henn}\ \emph {et~al.}(2009)\citenamefont {Henn},
  \citenamefont {Seman}, \citenamefont {Roati}, \citenamefont {Magalh\~aes},\
  and\ \citenamefont {Bagnato}}]{Henn2009}%
  \BibitemOpen
  \bibfield  {author} {\bibinfo {author} {\bibfnamefont {E.~A.~L.}\
  \bibnamefont {Henn}}, \bibinfo {author} {\bibfnamefont {J.~A.}\ \bibnamefont
  {Seman}}, \bibinfo {author} {\bibfnamefont {G.}~\bibnamefont {Roati}},
  \bibinfo {author} {\bibfnamefont {K.~M.~F.}\ \bibnamefont {Magalh\~aes}},\
  and\ \bibinfo {author} {\bibfnamefont {V.~S.}\ \bibnamefont {Bagnato}},\
  }\bibfield  {title} {\bibinfo {title} {Emergence of turbulence in an
  oscillating bose-einstein condensate},\ }\href
  {https://doi.org/10.1103/PhysRevLett.103.045301} {\bibfield  {journal}
  {\bibinfo  {journal} {Phys. Rev. Lett.}\ }\textbf {\bibinfo {volume} {103}},\
  \bibinfo {pages} {045301} (\bibinfo {year} {2009})}\BibitemShut {NoStop}%
\bibitem [{\citenamefont {Navon}\ \emph {et~al.}(2016)\citenamefont {Navon},
  \citenamefont {Gaunt}, \citenamefont {Smith},\ and\ \citenamefont
  {Hadzibabic}}]{Navon2016}%
  \BibitemOpen
  \bibfield  {author} {\bibinfo {author} {\bibfnamefont {N.}~\bibnamefont
  {Navon}}, \bibinfo {author} {\bibfnamefont {A.~L.}\ \bibnamefont {Gaunt}},
  \bibinfo {author} {\bibfnamefont {R.~P.}\ \bibnamefont {Smith}},\ and\
  \bibinfo {author} {\bibfnamefont {Z.}~\bibnamefont {Hadzibabic}},\ }\bibfield
   {title} {\bibinfo {title} {Emergence of a turbulent cascade in a quantum
  gas},\ }\href {https://doi.org/10.1038/nature20114} {\bibfield  {journal}
  {\bibinfo  {journal} {Nature}\ }\textbf {\bibinfo {volume} {539}},\ \bibinfo
  {pages} {72} (\bibinfo {year} {2016})}\BibitemShut {NoStop}%
\bibitem [{\citenamefont {Dogra}\ \emph {et~al.}(2023)\citenamefont {Dogra},
  \citenamefont {Martirosyan}, \citenamefont {Hilker}, \citenamefont {Glidden},
  \citenamefont {Etrych}, \citenamefont {Cao}, \citenamefont {Eigen},
  \citenamefont {Smith},\ and\ \citenamefont {Hadzibabic}}]{Dogra2023}%
  \BibitemOpen
  \bibfield  {author} {\bibinfo {author} {\bibfnamefont {L.~H.}\ \bibnamefont
  {Dogra}}, \bibinfo {author} {\bibfnamefont {G.}~\bibnamefont {Martirosyan}},
  \bibinfo {author} {\bibfnamefont {T.~A.}\ \bibnamefont {Hilker}}, \bibinfo
  {author} {\bibfnamefont {J.~A.~P.}\ \bibnamefont {Glidden}}, \bibinfo
  {author} {\bibfnamefont {J.}~\bibnamefont {Etrych}}, \bibinfo {author}
  {\bibfnamefont {A.}~\bibnamefont {Cao}}, \bibinfo {author} {\bibfnamefont
  {C.}~\bibnamefont {Eigen}}, \bibinfo {author} {\bibfnamefont {R.~P.}\
  \bibnamefont {Smith}},\ and\ \bibinfo {author} {\bibfnamefont
  {Z.}~\bibnamefont {Hadzibabic}},\ }\bibfield  {title} {\bibinfo {title}
  {Universal equation of state for wave turbulence in a quantum gas},\ }\href
  {https://doi.org/10.1038/s41586-023-06240-z} {\bibfield  {journal} {\bibinfo
  {journal} {Nature}\ }\textbf {\bibinfo {volume} {620}},\ \bibinfo {pages}
  {521} (\bibinfo {year} {2023})}\BibitemShut {NoStop}%
\bibitem [{\citenamefont {Cooper}\ \emph {et~al.}(2001)\citenamefont {Cooper},
  \citenamefont {Wilkin},\ and\ \citenamefont {Gunn}}]{Cooper2001}%
  \BibitemOpen
  \bibfield  {author} {\bibinfo {author} {\bibfnamefont {N.~R.}\ \bibnamefont
  {Cooper}}, \bibinfo {author} {\bibfnamefont {N.~K.}\ \bibnamefont {Wilkin}},\
  and\ \bibinfo {author} {\bibfnamefont {J.~M.~F.}\ \bibnamefont {Gunn}},\
  }\bibfield  {title} {\bibinfo {title} {Quantum phases of vortices in rotating
  bose-einstein condensates},\ }\href
  {https://doi.org/10.1103/PhysRevLett.87.120405} {\bibfield  {journal}
  {\bibinfo  {journal} {Phys. Rev. Lett.}\ }\textbf {\bibinfo {volume} {87}},\
  \bibinfo {pages} {120405} (\bibinfo {year} {2001})}\BibitemShut {NoStop}%
\bibitem [{coo(2008)}]{cooper2008rapidly}%
  \BibitemOpen
  \bibfield  {title} {\bibinfo {title} {Rapidly rotating atomic gases},\
  }\href@noop {} {\bibfield  {journal} {\bibinfo  {journal} {Advances in
  Physics}\ }\textbf {\bibinfo {volume} {57}},\ \bibinfo {pages} {539}
  (\bibinfo {year} {2008})}\BibitemShut {NoStop}%
\bibitem [{\citenamefont {Abo-Shaeer}\ \emph {et~al.}(2001)\citenamefont
  {Abo-Shaeer}, \citenamefont {Raman}, \citenamefont {Vogels},\ and\
  \citenamefont {Ketterle}}]{Abo2001}%
  \BibitemOpen
  \bibfield  {author} {\bibinfo {author} {\bibfnamefont {J.~R.}\ \bibnamefont
  {Abo-Shaeer}}, \bibinfo {author} {\bibfnamefont {C.}~\bibnamefont {Raman}},
  \bibinfo {author} {\bibfnamefont {J.~M.}\ \bibnamefont {Vogels}},\ and\
  \bibinfo {author} {\bibfnamefont {W.}~\bibnamefont {Ketterle}},\ }\bibfield
  {title} {\bibinfo {title} {Observation of vortex lattices in bose-einstein
  condensates},\ }\href {https://doi.org/10.1126/science.1060182} {\bibfield
  {journal} {\bibinfo  {journal} {Science}\ }\textbf {\bibinfo {volume}
  {292}},\ \bibinfo {pages} {476} (\bibinfo {year} {2001})},\ \Eprint
  {https://arxiv.org/abs/https://www.science.org/doi/pdf/10.1126/science.1060182}
  {https://www.science.org/doi/pdf/10.1126/science.1060182} \BibitemShut
  {NoStop}%
\bibitem [{\citenamefont {Mukherjee}\ \emph {et~al.}(2022)\citenamefont
  {Mukherjee}, \citenamefont {Shaffer}, \citenamefont {Patel}, \citenamefont
  {Yan}, \citenamefont {Wilson}, \citenamefont {Cr{\'e}pel}, \citenamefont
  {Fletcher},\ and\ \citenamefont {Zwierlein}}]{Mukherjee2022}%
  \BibitemOpen
  \bibfield  {author} {\bibinfo {author} {\bibfnamefont {B.}~\bibnamefont
  {Mukherjee}}, \bibinfo {author} {\bibfnamefont {A.}~\bibnamefont {Shaffer}},
  \bibinfo {author} {\bibfnamefont {P.~B.}\ \bibnamefont {Patel}}, \bibinfo
  {author} {\bibfnamefont {Z.}~\bibnamefont {Yan}}, \bibinfo {author}
  {\bibfnamefont {C.~C.}\ \bibnamefont {Wilson}}, \bibinfo {author}
  {\bibfnamefont {V.}~\bibnamefont {Cr{\'e}pel}}, \bibinfo {author}
  {\bibfnamefont {R.~J.}\ \bibnamefont {Fletcher}},\ and\ \bibinfo {author}
  {\bibfnamefont {M.}~\bibnamefont {Zwierlein}},\ }\bibfield  {title} {\bibinfo
  {title} {Crystallization of bosonic quantum hall states in a rotating quantum
  gas},\ }\href {https://doi.org/10.1038/s41586-021-04170-2} {\bibfield
  {journal} {\bibinfo  {journal} {Nature}\ }\textbf {\bibinfo {volume} {601}},\
  \bibinfo {pages} {58} (\bibinfo {year} {2022})}\BibitemShut {NoStop}%
\bibitem [{\citenamefont {L{\'e}onard}\ \emph {et~al.}(2023)\citenamefont
  {L{\'e}onard}, \citenamefont {Kim}, \citenamefont {Kwan}, \citenamefont
  {Segura}, \citenamefont {Grusdt}, \citenamefont {Repellin}, \citenamefont
  {Goldman},\ and\ \citenamefont {Greiner}}]{Leonard2023}%
  \BibitemOpen
  \bibfield  {author} {\bibinfo {author} {\bibfnamefont {J.}~\bibnamefont
  {L{\'e}onard}}, \bibinfo {author} {\bibfnamefont {S.}~\bibnamefont {Kim}},
  \bibinfo {author} {\bibfnamefont {J.}~\bibnamefont {Kwan}}, \bibinfo {author}
  {\bibfnamefont {P.}~\bibnamefont {Segura}}, \bibinfo {author} {\bibfnamefont
  {F.}~\bibnamefont {Grusdt}}, \bibinfo {author} {\bibfnamefont
  {C.}~\bibnamefont {Repellin}}, \bibinfo {author} {\bibfnamefont
  {N.}~\bibnamefont {Goldman}},\ and\ \bibinfo {author} {\bibfnamefont
  {M.}~\bibnamefont {Greiner}},\ }\bibfield  {title} {\bibinfo {title}
  {Realization of a fractional quantum hall state with ultracold atoms},\
  }\href {https://doi.org/10.1038/s41586-023-06122-4} {\bibfield  {journal}
  {\bibinfo  {journal} {Nature}\ }\textbf {\bibinfo {volume} {619}},\ \bibinfo
  {pages} {495} (\bibinfo {year} {2023})}\BibitemShut {NoStop}%
\bibitem [{\citenamefont {Lunt}\ \emph {et~al.}(2024)\citenamefont {Lunt},
  \citenamefont {Hill}, \citenamefont {Reiter}, \citenamefont {Preiss},
  \citenamefont {Ga\l{}ka},\ and\ \citenamefont {Jochim}}]{Lunt2024}%
  \BibitemOpen
  \bibfield  {author} {\bibinfo {author} {\bibfnamefont {P.}~\bibnamefont
  {Lunt}}, \bibinfo {author} {\bibfnamefont {P.}~\bibnamefont {Hill}}, \bibinfo
  {author} {\bibfnamefont {J.}~\bibnamefont {Reiter}}, \bibinfo {author}
  {\bibfnamefont {P.~M.}\ \bibnamefont {Preiss}}, \bibinfo {author}
  {\bibfnamefont {M.}~\bibnamefont {Ga\l{}ka}},\ and\ \bibinfo {author}
  {\bibfnamefont {S.}~\bibnamefont {Jochim}},\ }\bibfield  {title} {\bibinfo
  {title} {Realization of a laughlin state of two rapidly rotating fermions},\
  }\href {https://doi.org/10.1103/PhysRevLett.133.253401} {\bibfield  {journal}
  {\bibinfo  {journal} {Phys. Rev. Lett.}\ }\textbf {\bibinfo {volume} {133}},\
  \bibinfo {pages} {253401} (\bibinfo {year} {2024})}\BibitemShut {NoStop}%
\bibitem [{\citenamefont {Yao}\ \emph {et~al.}(2024)\citenamefont {Yao},
  \citenamefont {Chi}, \citenamefont {Mukherjee}, \citenamefont {Shaffer},
  \citenamefont {Zwierlein},\ and\ \citenamefont {Fletcher}}]{Yao2024}%
  \BibitemOpen
  \bibfield  {author} {\bibinfo {author} {\bibfnamefont {R.}~\bibnamefont
  {Yao}}, \bibinfo {author} {\bibfnamefont {S.}~\bibnamefont {Chi}}, \bibinfo
  {author} {\bibfnamefont {B.}~\bibnamefont {Mukherjee}}, \bibinfo {author}
  {\bibfnamefont {A.}~\bibnamefont {Shaffer}}, \bibinfo {author} {\bibfnamefont
  {M.}~\bibnamefont {Zwierlein}},\ and\ \bibinfo {author} {\bibfnamefont
  {R.~J.}\ \bibnamefont {Fletcher}},\ }\bibfield  {title} {\bibinfo {title}
  {Observation of chiral edge transport in a rapidly rotating quantum gas},\
  }\href {https://doi.org/10.1038/s41567-024-02617-7} {\bibfield  {journal}
  {\bibinfo  {journal} {Nature Physics}\ }\textbf {\bibinfo {volume} {20}},\
  \bibinfo {pages} {1726} (\bibinfo {year} {2024})}\BibitemShut {NoStop}%
\bibitem [{\citenamefont {Deng}\ \emph {et~al.}(2010)\citenamefont {Deng},
  \citenamefont {Haug},\ and\ \citenamefont {Yamamoto}}]{Deng2010}%
  \BibitemOpen
  \bibfield  {author} {\bibinfo {author} {\bibfnamefont {H.}~\bibnamefont
  {Deng}}, \bibinfo {author} {\bibfnamefont {H.}~\bibnamefont {Haug}},\ and\
  \bibinfo {author} {\bibfnamefont {Y.}~\bibnamefont {Yamamoto}},\ }\bibfield
  {title} {\bibinfo {title} {Exciton-polariton bose-einstein condensation},\
  }\href {https://doi.org/10.1103/RevModPhys.82.1489} {\bibfield  {journal}
  {\bibinfo  {journal} {Rev. Mod. Phys.}\ }\textbf {\bibinfo {volume} {82}},\
  \bibinfo {pages} {1489} (\bibinfo {year} {2010})}\BibitemShut {NoStop}%
\bibitem [{\citenamefont {Carusotto}\ and\ \citenamefont
  {Ciuti}(2013)}]{Carusotto2013}%
  \BibitemOpen
  \bibfield  {author} {\bibinfo {author} {\bibfnamefont {I.}~\bibnamefont
  {Carusotto}}\ and\ \bibinfo {author} {\bibfnamefont {C.}~\bibnamefont
  {Ciuti}},\ }\bibfield  {title} {\bibinfo {title} {Quantum fluids of light},\
  }\href {https://doi.org/10.1103/RevModPhys.85.299} {\bibfield  {journal}
  {\bibinfo  {journal} {Rev. Mod. Phys.}\ }\textbf {\bibinfo {volume} {85}},\
  \bibinfo {pages} {299} (\bibinfo {year} {2013})}\BibitemShut {NoStop}%
\bibitem [{\citenamefont {Keeling}\ and\ \citenamefont
  {Berloff}(2011)}]{Keeling01032011}%
  \BibitemOpen
  \bibfield  {author} {\bibinfo {author} {\bibfnamefont {J.}~\bibnamefont
  {Keeling}}\ and\ \bibinfo {author} {\bibfnamefont {N.~G.}\ \bibnamefont
  {Berloff}},\ }\bibfield  {title} {\bibinfo {title} {Exciton–polariton
  condensation},\ }\href {https://doi.org/10.1080/00107514.2010.550120}
  {\bibfield  {journal} {\bibinfo  {journal} {Contemporary Physics}\ }\textbf
  {\bibinfo {volume} {52}},\ \bibinfo {pages} {131} (\bibinfo {year} {2011})},\
  \Eprint {https://arxiv.org/abs/https://doi.org/10.1080/00107514.2010.550120}
  {https://doi.org/10.1080/00107514.2010.550120} \BibitemShut {NoStop}%
\bibitem [{\citenamefont {Byrnes}\ \emph {et~al.}(2014)\citenamefont {Byrnes},
  \citenamefont {Kim},\ and\ \citenamefont {Yamamoto}}]{Byrnes2014}%
  \BibitemOpen
  \bibfield  {author} {\bibinfo {author} {\bibfnamefont {T.}~\bibnamefont
  {Byrnes}}, \bibinfo {author} {\bibfnamefont {N.~Y.}\ \bibnamefont {Kim}},\
  and\ \bibinfo {author} {\bibfnamefont {Y.}~\bibnamefont {Yamamoto}},\
  }\bibfield  {title} {\bibinfo {title} {Exciton--polariton condensates},\
  }\href {https://doi.org/10.1038/nphys3143} {\bibfield  {journal} {\bibinfo
  {journal} {Nature Physics}\ }\textbf {\bibinfo {volume} {10}},\ \bibinfo
  {pages} {803} (\bibinfo {year} {2014})}\BibitemShut {NoStop}%
\bibitem [{\citenamefont {Guillet}\ and\ \citenamefont
  {Brimont}(2016)}]{guillet2016polariton}%
  \BibitemOpen
  \bibfield  {author} {\bibinfo {author} {\bibfnamefont {T.}~\bibnamefont
  {Guillet}}\ and\ \bibinfo {author} {\bibfnamefont {C.}~\bibnamefont
  {Brimont}},\ }\bibfield  {title} {\bibinfo {title} {Polariton condensates at
  room temperature},\ }\href@noop {} {\bibfield  {journal} {\bibinfo  {journal}
  {Comptes Rendus Physique}\ }\textbf {\bibinfo {volume} {17}},\ \bibinfo
  {pages} {946} (\bibinfo {year} {2016})}\BibitemShut {NoStop}%
\bibitem [{\citenamefont {Su}\ \emph {et~al.}(2020)\citenamefont {Su},
  \citenamefont {Ghosh}, \citenamefont {Wang}, \citenamefont {Liu},
  \citenamefont {Diederichs}, \citenamefont {Liew},\ and\ \citenamefont
  {Xiong}}]{Su2020}%
  \BibitemOpen
  \bibfield  {author} {\bibinfo {author} {\bibfnamefont {R.}~\bibnamefont
  {Su}}, \bibinfo {author} {\bibfnamefont {S.}~\bibnamefont {Ghosh}}, \bibinfo
  {author} {\bibfnamefont {J.}~\bibnamefont {Wang}}, \bibinfo {author}
  {\bibfnamefont {S.}~\bibnamefont {Liu}}, \bibinfo {author} {\bibfnamefont
  {C.}~\bibnamefont {Diederichs}}, \bibinfo {author} {\bibfnamefont {T.~C.~H.}\
  \bibnamefont {Liew}},\ and\ \bibinfo {author} {\bibfnamefont
  {Q.}~\bibnamefont {Xiong}},\ }\bibfield  {title} {\bibinfo {title}
  {Observation of exciton polariton condensation in a perovskite lattice at
  room temperature},\ }\href {https://doi.org/10.1038/s41567-019-0764-5}
  {\bibfield  {journal} {\bibinfo  {journal} {Nature Physics}\ }\textbf
  {\bibinfo {volume} {16}},\ \bibinfo {pages} {301} (\bibinfo {year}
  {2020})}\BibitemShut {NoStop}%
\bibitem [{\citenamefont {Lerario}\ \emph {et~al.}(2017)\citenamefont
  {Lerario}, \citenamefont {Fieramosca}, \citenamefont {Barachati},
  \citenamefont {Ballarini}, \citenamefont {Daskalakis}, \citenamefont
  {Dominici}, \citenamefont {De~Giorgi}, \citenamefont {Maier}, \citenamefont
  {Gigli}, \citenamefont {K{\'e}na-Cohen},\ and\ \citenamefont
  {Sanvitto}}]{Lerario2017}%
  \BibitemOpen
  \bibfield  {author} {\bibinfo {author} {\bibfnamefont {G.}~\bibnamefont
  {Lerario}}, \bibinfo {author} {\bibfnamefont {A.}~\bibnamefont {Fieramosca}},
  \bibinfo {author} {\bibfnamefont {F.}~\bibnamefont {Barachati}}, \bibinfo
  {author} {\bibfnamefont {D.}~\bibnamefont {Ballarini}}, \bibinfo {author}
  {\bibfnamefont {K.~S.}\ \bibnamefont {Daskalakis}}, \bibinfo {author}
  {\bibfnamefont {L.}~\bibnamefont {Dominici}}, \bibinfo {author}
  {\bibfnamefont {M.}~\bibnamefont {De~Giorgi}}, \bibinfo {author}
  {\bibfnamefont {S.~A.}\ \bibnamefont {Maier}}, \bibinfo {author}
  {\bibfnamefont {G.}~\bibnamefont {Gigli}}, \bibinfo {author} {\bibfnamefont
  {S.}~\bibnamefont {K{\'e}na-Cohen}},\ and\ \bibinfo {author} {\bibfnamefont
  {D.}~\bibnamefont {Sanvitto}},\ }\bibfield  {title} {\bibinfo {title}
  {Room-temperature superfluidity in a polariton condensate},\ }\href
  {https://doi.org/10.1038/nphys4147} {\bibfield  {journal} {\bibinfo
  {journal} {Nature Physics}\ }\textbf {\bibinfo {volume} {13}},\ \bibinfo
  {pages} {837} (\bibinfo {year} {2017})}\BibitemShut {NoStop}%
\bibitem [{\citenamefont {Peng}\ \emph {et~al.}(2022)\citenamefont {Peng},
  \citenamefont {Tao}, \citenamefont {Haeberl{\'e}}, \citenamefont {Li},
  \citenamefont {Jin}, \citenamefont {Fleming}, \citenamefont {K{\'e}na-Cohen},
  \citenamefont {Zhang},\ and\ \citenamefont {Bao}}]{Peng2022}%
  \BibitemOpen
  \bibfield  {author} {\bibinfo {author} {\bibfnamefont {K.}~\bibnamefont
  {Peng}}, \bibinfo {author} {\bibfnamefont {R.}~\bibnamefont {Tao}}, \bibinfo
  {author} {\bibfnamefont {L.}~\bibnamefont {Haeberl{\'e}}}, \bibinfo {author}
  {\bibfnamefont {Q.}~\bibnamefont {Li}}, \bibinfo {author} {\bibfnamefont
  {D.}~\bibnamefont {Jin}}, \bibinfo {author} {\bibfnamefont {G.~R.}\
  \bibnamefont {Fleming}}, \bibinfo {author} {\bibfnamefont {S.}~\bibnamefont
  {K{\'e}na-Cohen}}, \bibinfo {author} {\bibfnamefont {X.}~\bibnamefont
  {Zhang}},\ and\ \bibinfo {author} {\bibfnamefont {W.}~\bibnamefont {Bao}},\
  }\bibfield  {title} {\bibinfo {title} {Room-temperature polariton quantum
  fluids in halide perovskites},\ }\href
  {https://doi.org/10.1038/s41467-022-34987-y} {\bibfield  {journal} {\bibinfo
  {journal} {Nature Communications}\ }\textbf {\bibinfo {volume} {13}},\
  \bibinfo {pages} {7388} (\bibinfo {year} {2022})}\BibitemShut {NoStop}%
\bibitem [{\citenamefont {Lagoudakis}\ \emph {et~al.}(2009)\citenamefont
  {Lagoudakis}, \citenamefont {Ostatnický}, \citenamefont {Kavokin},
  \citenamefont {Rubo}, \citenamefont {André},\ and\ \citenamefont
  {Deveaud-Plédran}}]{LagoudakisHalfQuatumVortices}%
  \BibitemOpen
  \bibfield  {author} {\bibinfo {author} {\bibfnamefont {K.~G.}\ \bibnamefont
  {Lagoudakis}}, \bibinfo {author} {\bibfnamefont {T.}~\bibnamefont
  {Ostatnický}}, \bibinfo {author} {\bibfnamefont {A.~V.}\ \bibnamefont
  {Kavokin}}, \bibinfo {author} {\bibfnamefont {Y.~G.}\ \bibnamefont {Rubo}},
  \bibinfo {author} {\bibfnamefont {R.}~\bibnamefont {André}},\ and\ \bibinfo
  {author} {\bibfnamefont {B.}~\bibnamefont {Deveaud-Plédran}},\ }\bibfield
  {title} {\bibinfo {title} {Observation of half-quantum vortices in an
  exciton-polariton condensate},\ }\href
  {https://doi.org/10.1126/science.1177980} {\bibfield  {journal} {\bibinfo
  {journal} {Science}\ }\textbf {\bibinfo {volume} {326}},\ \bibinfo {pages}
  {974} (\bibinfo {year} {2009})},\ \Eprint
  {https://arxiv.org/abs/https://www.science.org/doi/pdf/10.1126/science.1177980}
  {https://www.science.org/doi/pdf/10.1126/science.1177980} \BibitemShut
  {NoStop}%
\bibitem [{\citenamefont {Lagoudakis}\ \emph {et~al.}(2008)\citenamefont
  {Lagoudakis}, \citenamefont {Wouters}, \citenamefont {Richard}, \citenamefont
  {Baas}, \citenamefont {Carusotto}, \citenamefont {Andr{\'e}}, \citenamefont
  {Dang},\ and\ \citenamefont {Deveaud-Pl{\'e}dran}}]{Lagoudakis2008}%
  \BibitemOpen
  \bibfield  {author} {\bibinfo {author} {\bibfnamefont {K.~G.}\ \bibnamefont
  {Lagoudakis}}, \bibinfo {author} {\bibfnamefont {M.}~\bibnamefont {Wouters}},
  \bibinfo {author} {\bibfnamefont {M.}~\bibnamefont {Richard}}, \bibinfo
  {author} {\bibfnamefont {A.}~\bibnamefont {Baas}}, \bibinfo {author}
  {\bibfnamefont {I.}~\bibnamefont {Carusotto}}, \bibinfo {author}
  {\bibfnamefont {R.}~\bibnamefont {Andr{\'e}}}, \bibinfo {author}
  {\bibfnamefont {L.~S.}\ \bibnamefont {Dang}},\ and\ \bibinfo {author}
  {\bibfnamefont {B.}~\bibnamefont {Deveaud-Pl{\'e}dran}},\ }\bibfield  {title}
  {\bibinfo {title} {Quantized vortices in an exciton--polariton condensate},\
  }\href {https://doi.org/10.1038/nphys1051} {\bibfield  {journal} {\bibinfo
  {journal} {Nature Physics}\ }\textbf {\bibinfo {volume} {4}},\ \bibinfo
  {pages} {706} (\bibinfo {year} {2008})}\BibitemShut {NoStop}%
\bibitem [{\citenamefont {Sanvitto}\ \emph {et~al.}(2010)\citenamefont
  {Sanvitto}, \citenamefont {Marchetti}, \citenamefont {Szyma{\'n}ska},
  \citenamefont {Tosi}, \citenamefont {Baudisch}, \citenamefont {Laussy},
  \citenamefont {Krizhanovskii}, \citenamefont {Skolnick}, \citenamefont
  {Marrucci}, \citenamefont {Lema{\^\i}tre}, \citenamefont {Bloch},
  \citenamefont {Tejedor},\ and\ \citenamefont {Vi{\~n}a}}]{Sanvitto2010}%
  \BibitemOpen
  \bibfield  {author} {\bibinfo {author} {\bibfnamefont {D.}~\bibnamefont
  {Sanvitto}}, \bibinfo {author} {\bibfnamefont {F.~M.}\ \bibnamefont
  {Marchetti}}, \bibinfo {author} {\bibfnamefont {M.~H.}\ \bibnamefont
  {Szyma{\'n}ska}}, \bibinfo {author} {\bibfnamefont {G.}~\bibnamefont {Tosi}},
  \bibinfo {author} {\bibfnamefont {M.}~\bibnamefont {Baudisch}}, \bibinfo
  {author} {\bibfnamefont {F.~P.}\ \bibnamefont {Laussy}}, \bibinfo {author}
  {\bibfnamefont {D.~N.}\ \bibnamefont {Krizhanovskii}}, \bibinfo {author}
  {\bibfnamefont {M.~S.}\ \bibnamefont {Skolnick}}, \bibinfo {author}
  {\bibfnamefont {L.}~\bibnamefont {Marrucci}}, \bibinfo {author}
  {\bibfnamefont {A.}~\bibnamefont {Lema{\^\i}tre}}, \bibinfo {author}
  {\bibfnamefont {J.}~\bibnamefont {Bloch}}, \bibinfo {author} {\bibfnamefont
  {C.}~\bibnamefont {Tejedor}},\ and\ \bibinfo {author} {\bibfnamefont
  {L.}~\bibnamefont {Vi{\~n}a}},\ }\bibfield  {title} {\bibinfo {title}
  {Persistent currents and quantized vortices in a polariton superfluid},\
  }\href {https://doi.org/10.1038/nphys1668} {\bibfield  {journal} {\bibinfo
  {journal} {Nature Physics}\ }\textbf {\bibinfo {volume} {6}},\ \bibinfo
  {pages} {527} (\bibinfo {year} {2010})}\BibitemShut {NoStop}%
\bibitem [{\citenamefont {Lagoudakis}\ \emph {et~al.}(2011)\citenamefont
  {Lagoudakis}, \citenamefont {Manni}, \citenamefont {Pietka}, \citenamefont
  {Wouters}, \citenamefont {Liew}, \citenamefont {Savona}, \citenamefont
  {Kavokin}, \citenamefont {Andr\'e},\ and\ \citenamefont
  {Deveaud-Pl\'edran}}]{Lagoudakis2011}%
  \BibitemOpen
  \bibfield  {author} {\bibinfo {author} {\bibfnamefont {K.~G.}\ \bibnamefont
  {Lagoudakis}}, \bibinfo {author} {\bibfnamefont {F.}~\bibnamefont {Manni}},
  \bibinfo {author} {\bibfnamefont {B.}~\bibnamefont {Pietka}}, \bibinfo
  {author} {\bibfnamefont {M.}~\bibnamefont {Wouters}}, \bibinfo {author}
  {\bibfnamefont {T.~C.~H.}\ \bibnamefont {Liew}}, \bibinfo {author}
  {\bibfnamefont {V.}~\bibnamefont {Savona}}, \bibinfo {author} {\bibfnamefont
  {A.~V.}\ \bibnamefont {Kavokin}}, \bibinfo {author} {\bibfnamefont
  {R.}~\bibnamefont {Andr\'e}},\ and\ \bibinfo {author} {\bibfnamefont
  {B.}~\bibnamefont {Deveaud-Pl\'edran}},\ }\bibfield  {title} {\bibinfo
  {title} {Probing the dynamics of spontaneous quantum vortices in polariton
  superfluids},\ }\href {https://doi.org/10.1103/PhysRevLett.106.115301}
  {\bibfield  {journal} {\bibinfo  {journal} {Phys. Rev. Lett.}\ }\textbf
  {\bibinfo {volume} {106}},\ \bibinfo {pages} {115301} (\bibinfo {year}
  {2011})}\BibitemShut {NoStop}%
\bibitem [{\citenamefont {Nardin}\ \emph {et~al.}(2011)\citenamefont {Nardin},
  \citenamefont {Grosso}, \citenamefont {L{\'e}ger}, \citenamefont {Pi{\c
  e}tka}, \citenamefont {Morier-Genoud},\ and\ \citenamefont
  {Deveaud-Pl{\'e}dran}}]{Nardin2011}%
  \BibitemOpen
  \bibfield  {author} {\bibinfo {author} {\bibfnamefont {G.}~\bibnamefont
  {Nardin}}, \bibinfo {author} {\bibfnamefont {G.}~\bibnamefont {Grosso}},
  \bibinfo {author} {\bibfnamefont {Y.}~\bibnamefont {L{\'e}ger}}, \bibinfo
  {author} {\bibfnamefont {B.}~\bibnamefont {Pi{\c e}tka}}, \bibinfo {author}
  {\bibfnamefont {F.}~\bibnamefont {Morier-Genoud}},\ and\ \bibinfo {author}
  {\bibfnamefont {B.}~\bibnamefont {Deveaud-Pl{\'e}dran}},\ }\bibfield  {title}
  {\bibinfo {title} {Hydrodynamic nucleation of quantized vortex pairs in a
  polariton quantum fluid},\ }\href {https://doi.org/10.1038/nphys1959}
  {\bibfield  {journal} {\bibinfo  {journal} {Nature Physics}\ }\textbf
  {\bibinfo {volume} {7}},\ \bibinfo {pages} {635} (\bibinfo {year}
  {2011})}\BibitemShut {NoStop}%
\bibitem [{\citenamefont {Sanvitto}\ and\ \citenamefont
  {K{\'e}na-Cohen}(2016)}]{Sanvitto2016}%
  \BibitemOpen
  \bibfield  {author} {\bibinfo {author} {\bibfnamefont {D.}~\bibnamefont
  {Sanvitto}}\ and\ \bibinfo {author} {\bibfnamefont {S.}~\bibnamefont
  {K{\'e}na-Cohen}},\ }\bibfield  {title} {\bibinfo {title} {The road towards
  polaritonic devices},\ }\href {https://doi.org/10.1038/nmat4668} {\bibfield
  {journal} {\bibinfo  {journal} {Nature Materials}\ }\textbf {\bibinfo
  {volume} {15}},\ \bibinfo {pages} {1061} (\bibinfo {year}
  {2016})}\BibitemShut {NoStop}%
\bibitem [{\citenamefont {Tosi}\ \emph {et~al.}(2012)\citenamefont {Tosi},
  \citenamefont {Christmann}, \citenamefont {Berloff}, \citenamefont {Tsotsis},
  \citenamefont {Gao}, \citenamefont {Hatzopoulos}, \citenamefont {Savvidis},\
  and\ \citenamefont {Baumberg}}]{Tosi2012}%
  \BibitemOpen
  \bibfield  {author} {\bibinfo {author} {\bibfnamefont {G.}~\bibnamefont
  {Tosi}}, \bibinfo {author} {\bibfnamefont {G.}~\bibnamefont {Christmann}},
  \bibinfo {author} {\bibfnamefont {N.~G.}\ \bibnamefont {Berloff}}, \bibinfo
  {author} {\bibfnamefont {P.}~\bibnamefont {Tsotsis}}, \bibinfo {author}
  {\bibfnamefont {T.}~\bibnamefont {Gao}}, \bibinfo {author} {\bibfnamefont
  {Z.}~\bibnamefont {Hatzopoulos}}, \bibinfo {author} {\bibfnamefont {P.~G.}\
  \bibnamefont {Savvidis}},\ and\ \bibinfo {author} {\bibfnamefont {J.~J.}\
  \bibnamefont {Baumberg}},\ }\bibfield  {title} {\bibinfo {title}
  {Geometrically locked vortex lattices in semiconductor quantum fluids},\
  }\href {https://doi.org/10.1038/ncomms2255} {\bibfield  {journal} {\bibinfo
  {journal} {Nature Communications}\ }\textbf {\bibinfo {volume} {3}},\
  \bibinfo {pages} {1243} (\bibinfo {year} {2012})}\BibitemShut {NoStop}%
\bibitem [{\citenamefont {Zasedatelev}\ \emph {et~al.}(2019)\citenamefont
  {Zasedatelev}, \citenamefont {Baranikov}, \citenamefont {Urbonas},
  \citenamefont {Scafirimuto}, \citenamefont {Scherf}, \citenamefont
  {St{\"o}ferle}, \citenamefont {Mahrt},\ and\ \citenamefont
  {Lagoudakis}}]{Zasedatelev2019}%
  \BibitemOpen
  \bibfield  {author} {\bibinfo {author} {\bibfnamefont {A.~V.}\ \bibnamefont
  {Zasedatelev}}, \bibinfo {author} {\bibfnamefont {A.~V.}\ \bibnamefont
  {Baranikov}}, \bibinfo {author} {\bibfnamefont {D.}~\bibnamefont {Urbonas}},
  \bibinfo {author} {\bibfnamefont {F.}~\bibnamefont {Scafirimuto}}, \bibinfo
  {author} {\bibfnamefont {U.}~\bibnamefont {Scherf}}, \bibinfo {author}
  {\bibfnamefont {T.}~\bibnamefont {St{\"o}ferle}}, \bibinfo {author}
  {\bibfnamefont {R.~F.}\ \bibnamefont {Mahrt}},\ and\ \bibinfo {author}
  {\bibfnamefont {P.~G.}\ \bibnamefont {Lagoudakis}},\ }\bibfield  {title}
  {\bibinfo {title} {A room-temperature organic polariton transistor},\ }\href
  {https://doi.org/10.1038/s41566-019-0392-8} {\bibfield  {journal} {\bibinfo
  {journal} {Nature Photonics}\ }\textbf {\bibinfo {volume} {13}},\ \bibinfo
  {pages} {378} (\bibinfo {year} {2019})}\BibitemShut {NoStop}%
\bibitem [{\citenamefont {Sun}\ \emph {et~al.}(2017)\citenamefont {Sun},
  \citenamefont {Yoon}, \citenamefont {Steger}, \citenamefont {Liu},
  \citenamefont {Pfeiffer}, \citenamefont {West}, \citenamefont {Snoke},\ and\
  \citenamefont {Nelson}}]{Sun2017}%
  \BibitemOpen
  \bibfield  {author} {\bibinfo {author} {\bibfnamefont {Y.}~\bibnamefont
  {Sun}}, \bibinfo {author} {\bibfnamefont {Y.}~\bibnamefont {Yoon}}, \bibinfo
  {author} {\bibfnamefont {M.}~\bibnamefont {Steger}}, \bibinfo {author}
  {\bibfnamefont {G.}~\bibnamefont {Liu}}, \bibinfo {author} {\bibfnamefont
  {L.~N.}\ \bibnamefont {Pfeiffer}}, \bibinfo {author} {\bibfnamefont
  {K.}~\bibnamefont {West}}, \bibinfo {author} {\bibfnamefont {D.~W.}\
  \bibnamefont {Snoke}},\ and\ \bibinfo {author} {\bibfnamefont {K.~A.}\
  \bibnamefont {Nelson}},\ }\bibfield  {title} {\bibinfo {title} {Direct
  measurement of polariton--polariton interaction strength},\ }\href
  {https://doi.org/10.1038/nphys4148} {\bibfield  {journal} {\bibinfo
  {journal} {Nature Physics}\ }\textbf {\bibinfo {volume} {13}},\ \bibinfo
  {pages} {870} (\bibinfo {year} {2017})}\BibitemShut {NoStop}%
\bibitem [{\citenamefont {Takemura}\ \emph {et~al.}(2014)\citenamefont
  {Takemura}, \citenamefont {Trebaol}, \citenamefont {Wouters}, \citenamefont
  {Portella-Oberli},\ and\ \citenamefont {Deveaud}}]{Takemura2014}%
  \BibitemOpen
  \bibfield  {author} {\bibinfo {author} {\bibfnamefont {N.}~\bibnamefont
  {Takemura}}, \bibinfo {author} {\bibfnamefont {S.}~\bibnamefont {Trebaol}},
  \bibinfo {author} {\bibfnamefont {M.}~\bibnamefont {Wouters}}, \bibinfo
  {author} {\bibfnamefont {M.~T.}\ \bibnamefont {Portella-Oberli}},\ and\
  \bibinfo {author} {\bibfnamefont {B.}~\bibnamefont {Deveaud}},\ }\bibfield
  {title} {\bibinfo {title} {Polaritonic feshbach resonance},\ }\href
  {https://doi.org/10.1038/nphys2999} {\bibfield  {journal} {\bibinfo
  {journal} {Nature Physics}\ }\textbf {\bibinfo {volume} {10}},\ \bibinfo
  {pages} {500} (\bibinfo {year} {2014})}\BibitemShut {NoStop}%
\bibitem [{\citenamefont {Berger}\ \emph {et~al.}(2020)\citenamefont {Berger},
  \citenamefont {Schmidt}, \citenamefont {Ma}, \citenamefont {Schumacher},
  \citenamefont {Schneider}, \citenamefont {H\"ofling},\ and\ \citenamefont
  {A\ss{}mann}}]{Berger2020}%
  \BibitemOpen
  \bibfield  {author} {\bibinfo {author} {\bibfnamefont {B.}~\bibnamefont
  {Berger}}, \bibinfo {author} {\bibfnamefont {D.}~\bibnamefont {Schmidt}},
  \bibinfo {author} {\bibfnamefont {X.}~\bibnamefont {Ma}}, \bibinfo {author}
  {\bibfnamefont {S.}~\bibnamefont {Schumacher}}, \bibinfo {author}
  {\bibfnamefont {C.}~\bibnamefont {Schneider}}, \bibinfo {author}
  {\bibfnamefont {S.}~\bibnamefont {H\"ofling}},\ and\ \bibinfo {author}
  {\bibfnamefont {M.}~\bibnamefont {A\ss{}mann}},\ }\bibfield  {title}
  {\bibinfo {title} {Formation dynamics of exciton-polariton vortices created
  by nonresonant annular pumping},\ }\href
  {https://doi.org/10.1103/PhysRevB.101.245309} {\bibfield  {journal} {\bibinfo
   {journal} {Phys. Rev. B}\ }\textbf {\bibinfo {volume} {101}},\ \bibinfo
  {pages} {245309} (\bibinfo {year} {2020})}\BibitemShut {NoStop}%
\bibitem [{\citenamefont {Panico}\ \emph {et~al.}(2023)\citenamefont {Panico},
  \citenamefont {Comaron}, \citenamefont {Matuszewski}, \citenamefont
  {Lanotte}, \citenamefont {Trypogeorgos}, \citenamefont {Gigli}, \citenamefont
  {Giorgi}, \citenamefont {Ardizzone}, \citenamefont {Sanvitto},\ and\
  \citenamefont {Ballarini}}]{Panico2023}%
  \BibitemOpen
  \bibfield  {author} {\bibinfo {author} {\bibfnamefont {R.}~\bibnamefont
  {Panico}}, \bibinfo {author} {\bibfnamefont {P.}~\bibnamefont {Comaron}},
  \bibinfo {author} {\bibfnamefont {M.}~\bibnamefont {Matuszewski}}, \bibinfo
  {author} {\bibfnamefont {A.~S.}\ \bibnamefont {Lanotte}}, \bibinfo {author}
  {\bibfnamefont {D.}~\bibnamefont {Trypogeorgos}}, \bibinfo {author}
  {\bibfnamefont {G.}~\bibnamefont {Gigli}}, \bibinfo {author} {\bibfnamefont
  {M.~D.}\ \bibnamefont {Giorgi}}, \bibinfo {author} {\bibfnamefont
  {V.}~\bibnamefont {Ardizzone}}, \bibinfo {author} {\bibfnamefont
  {D.}~\bibnamefont {Sanvitto}},\ and\ \bibinfo {author} {\bibfnamefont
  {D.}~\bibnamefont {Ballarini}},\ }\bibfield  {title} {\bibinfo {title} {Onset
  of vortex clustering and inverse energy cascade in dissipative quantum
  fluids},\ }\href {https://doi.org/10.1038/s41566-023-01174-4} {\bibfield
  {journal} {\bibinfo  {journal} {Nature Photonics}\ }\textbf {\bibinfo
  {volume} {17}},\ \bibinfo {pages} {451} (\bibinfo {year} {2023})}\BibitemShut
  {NoStop}%
\bibitem [{\citenamefont {Comaron}\ \emph {et~al.}(2025)\citenamefont
  {Comaron}, \citenamefont {Panico}, \citenamefont {Ballarini},\ and\
  \citenamefont {Matuszewski}}]{Comaron2025}%
  \BibitemOpen
  \bibfield  {author} {\bibinfo {author} {\bibfnamefont {P.}~\bibnamefont
  {Comaron}}, \bibinfo {author} {\bibfnamefont {R.}~\bibnamefont {Panico}},
  \bibinfo {author} {\bibfnamefont {D.}~\bibnamefont {Ballarini}},\ and\
  \bibinfo {author} {\bibfnamefont {M.}~\bibnamefont {Matuszewski}},\
  }\bibfield  {title} {\bibinfo {title} {Dynamics of onsager vortex clustering
  in decaying turbulent polariton quantum fluids},\ }\href
  {https://doi.org/10.1103/PhysRevResearch.7.L022006} {\bibfield  {journal}
  {\bibinfo  {journal} {Phys. Rev. Res.}\ }\textbf {\bibinfo {volume} {7}},\
  \bibinfo {pages} {L022006} (\bibinfo {year} {2025})}\BibitemShut {NoStop}%
\bibitem [{\citenamefont {Drori}\ \emph {et~al.}(2023)\citenamefont {Drori},
  \citenamefont {Das}, \citenamefont {Zohar}, \citenamefont {Winer},
  \citenamefont {Poem}, \citenamefont {Poddubny},\ and\ \citenamefont
  {Firstenberg}}]{Drori2023}%
  \BibitemOpen
  \bibfield  {author} {\bibinfo {author} {\bibfnamefont {L.}~\bibnamefont
  {Drori}}, \bibinfo {author} {\bibfnamefont {B.~C.}\ \bibnamefont {Das}},
  \bibinfo {author} {\bibfnamefont {T.~D.}\ \bibnamefont {Zohar}}, \bibinfo
  {author} {\bibfnamefont {G.}~\bibnamefont {Winer}}, \bibinfo {author}
  {\bibfnamefont {E.}~\bibnamefont {Poem}}, \bibinfo {author} {\bibfnamefont
  {A.}~\bibnamefont {Poddubny}},\ and\ \bibinfo {author} {\bibfnamefont
  {O.}~\bibnamefont {Firstenberg}},\ }\bibfield  {title} {\bibinfo {title}
  {Quantum vortices of strongly interacting photons},\ }\href
  {https://doi.org/10.1126/science.adh5315} {\bibfield  {journal} {\bibinfo
  {journal} {Science}\ }\textbf {\bibinfo {volume} {381}},\ \bibinfo {pages}
  {193} (\bibinfo {year} {2023})}\BibitemShut {NoStop}%
\bibitem [{\citenamefont {Huang}\ \emph {et~al.}(2025)\citenamefont {Huang},
  \citenamefont {Mao}, \citenamefont {Li}, \citenamefont {Yuan}, \citenamefont
  {Zheng}, \citenamefont {Zhai}, \citenamefont {Dai}, \citenamefont {Fu},
  \citenamefont {Bao}, \citenamefont {Yang}, \citenamefont {Dai}, \citenamefont
  {Li}, \citenamefont {Gong},\ and\ \citenamefont {Wang}}]{Huang2025}%
  \BibitemOpen
  \bibfield  {author} {\bibinfo {author} {\bibfnamefont {J.}~\bibnamefont
  {Huang}}, \bibinfo {author} {\bibfnamefont {J.}~\bibnamefont {Mao}}, \bibinfo
  {author} {\bibfnamefont {X.}~\bibnamefont {Li}}, \bibinfo {author}
  {\bibfnamefont {J.}~\bibnamefont {Yuan}}, \bibinfo {author} {\bibfnamefont
  {Y.}~\bibnamefont {Zheng}}, \bibinfo {author} {\bibfnamefont
  {C.}~\bibnamefont {Zhai}}, \bibinfo {author} {\bibfnamefont {T.}~\bibnamefont
  {Dai}}, \bibinfo {author} {\bibfnamefont {Z.}~\bibnamefont {Fu}}, \bibinfo
  {author} {\bibfnamefont {J.}~\bibnamefont {Bao}}, \bibinfo {author}
  {\bibfnamefont {Y.}~\bibnamefont {Yang}}, \bibinfo {author} {\bibfnamefont
  {D.}~\bibnamefont {Dai}}, \bibinfo {author} {\bibfnamefont {Y.}~\bibnamefont
  {Li}}, \bibinfo {author} {\bibfnamefont {Q.}~\bibnamefont {Gong}},\ and\
  \bibinfo {author} {\bibfnamefont {J.}~\bibnamefont {Wang}},\ }\bibfield
  {title} {\bibinfo {title} {Integrated optical entangled quantum vortex
  emitters},\ }\href {https://doi.org/10.1038/s41566-025-01620-5} {\bibfield
  {journal} {\bibinfo  {journal} {Nature Photonics}\ }\textbf {\bibinfo
  {volume} {19}},\ \bibinfo {pages} {471} (\bibinfo {year} {2025})}\BibitemShut
  {NoStop}%
\bibitem [{\citenamefont {Swarnkar}\ \emph {et~al.}(2016)\citenamefont
  {Swarnkar}, \citenamefont {Marshall}, \citenamefont {Sanehira}, \citenamefont
  {Chernomordik}, \citenamefont {Moore}, \citenamefont {Christians},
  \citenamefont {Chakrabarti},\ and\ \citenamefont {Luther}}]{Swarnkar2016}%
  \BibitemOpen
  \bibfield  {author} {\bibinfo {author} {\bibfnamefont {A.}~\bibnamefont
  {Swarnkar}}, \bibinfo {author} {\bibfnamefont {A.~R.}\ \bibnamefont
  {Marshall}}, \bibinfo {author} {\bibfnamefont {E.~M.}\ \bibnamefont
  {Sanehira}}, \bibinfo {author} {\bibfnamefont {B.~D.}\ \bibnamefont
  {Chernomordik}}, \bibinfo {author} {\bibfnamefont {D.~T.}\ \bibnamefont
  {Moore}}, \bibinfo {author} {\bibfnamefont {J.~A.}\ \bibnamefont
  {Christians}}, \bibinfo {author} {\bibfnamefont {T.}~\bibnamefont
  {Chakrabarti}},\ and\ \bibinfo {author} {\bibfnamefont {J.~M.}\ \bibnamefont
  {Luther}},\ }\bibfield  {title} {\bibinfo {title} {Quantum
  dot\&\#x2013;induced phase stabilization of \&\#x3b1;-cspbi<sub>3</sub>
  perovskite for high-efficiency photovoltaics},\ }\href
  {https://doi.org/10.1126/science.aag2700} {\bibfield  {journal} {\bibinfo
  {journal} {Science}\ }\textbf {\bibinfo {volume} {354}},\ \bibinfo {pages}
  {92} (\bibinfo {year} {2016})},\ \Eprint
  {https://arxiv.org/abs/https://www.science.org/doi/pdf/10.1126/science.aag2700}
  {https://www.science.org/doi/pdf/10.1126/science.aag2700} \BibitemShut
  {NoStop}%
\bibitem [{\citenamefont {Shamsi}\ \emph {et~al.}(2019)\citenamefont {Shamsi},
  \citenamefont {Urban}, \citenamefont {Imran}, \citenamefont {De~Trizio},\
  and\ \citenamefont {Manna}}]{Shamsi2019}%
  \BibitemOpen
  \bibfield  {author} {\bibinfo {author} {\bibfnamefont {J.}~\bibnamefont
  {Shamsi}}, \bibinfo {author} {\bibfnamefont {A.~S.}\ \bibnamefont {Urban}},
  \bibinfo {author} {\bibfnamefont {M.}~\bibnamefont {Imran}}, \bibinfo
  {author} {\bibfnamefont {L.}~\bibnamefont {De~Trizio}},\ and\ \bibinfo
  {author} {\bibfnamefont {L.}~\bibnamefont {Manna}},\ }\bibfield  {title}
  {\bibinfo {title} {Metal halide perovskite nanocrystals: Synthesis,
  post-synthesis modifications, and their optical properties},\ }\href
  {https://doi.org/10.1021/acs.chemrev.8b00644} {\bibfield  {journal} {\bibinfo
   {journal} {Chemical Reviews}\ }\textbf {\bibinfo {volume} {119}},\ \bibinfo
  {pages} {3296} (\bibinfo {year} {2019})}\BibitemShut {NoStop}%
\bibitem [{\citenamefont {Yu}\ \emph {et~al.}(2020)\citenamefont {Yu},
  \citenamefont {Liu}, \citenamefont {Chen}, \citenamefont {Li}, \citenamefont
  {Xu}, \citenamefont {Wang}, \citenamefont {Zhao},\ and\ \citenamefont
  {Zhang}}]{yu2020perovskite}%
  \BibitemOpen
  \bibfield  {author} {\bibinfo {author} {\bibfnamefont {J.}~\bibnamefont
  {Yu}}, \bibinfo {author} {\bibfnamefont {G.}~\bibnamefont {Liu}}, \bibinfo
  {author} {\bibfnamefont {C.}~\bibnamefont {Chen}}, \bibinfo {author}
  {\bibfnamefont {Y.}~\bibnamefont {Li}}, \bibinfo {author} {\bibfnamefont
  {M.}~\bibnamefont {Xu}}, \bibinfo {author} {\bibfnamefont {T.}~\bibnamefont
  {Wang}}, \bibinfo {author} {\bibfnamefont {G.}~\bibnamefont {Zhao}},\ and\
  \bibinfo {author} {\bibfnamefont {L.}~\bibnamefont {Zhang}},\ }\bibfield
  {title} {\bibinfo {title} {Perovskite cspbbr 3 crystals: growth and
  applications},\ }\href@noop {} {\bibfield  {journal} {\bibinfo  {journal}
  {Journal of Materials Chemistry C}\ }\textbf {\bibinfo {volume} {8}},\
  \bibinfo {pages} {6326} (\bibinfo {year} {2020})}\BibitemShut {NoStop}%
\bibitem [{\citenamefont {Protesescu}\ \emph {et~al.}(2015)\citenamefont
  {Protesescu}, \citenamefont {Yakunin}, \citenamefont {Bodnarchuk},
  \citenamefont {Krieg}, \citenamefont {Caputo}, \citenamefont {Hendon},
  \citenamefont {Yang}, \citenamefont {Walsh},\ and\ \citenamefont
  {Kovalenko}}]{Protesescu2015}%
  \BibitemOpen
  \bibfield  {author} {\bibinfo {author} {\bibfnamefont {L.}~\bibnamefont
  {Protesescu}}, \bibinfo {author} {\bibfnamefont {S.}~\bibnamefont {Yakunin}},
  \bibinfo {author} {\bibfnamefont {M.~I.}\ \bibnamefont {Bodnarchuk}},
  \bibinfo {author} {\bibfnamefont {F.}~\bibnamefont {Krieg}}, \bibinfo
  {author} {\bibfnamefont {R.}~\bibnamefont {Caputo}}, \bibinfo {author}
  {\bibfnamefont {C.~H.}\ \bibnamefont {Hendon}}, \bibinfo {author}
  {\bibfnamefont {R.~X.}\ \bibnamefont {Yang}}, \bibinfo {author}
  {\bibfnamefont {A.}~\bibnamefont {Walsh}},\ and\ \bibinfo {author}
  {\bibfnamefont {M.~V.}\ \bibnamefont {Kovalenko}},\ }\bibfield  {title}
  {\bibinfo {title} {Nanocrystals of cesium lead halide perovskites (cspbx3, x
  = cl, br, and i): Novel optoelectronic materials showing bright emission with
  wide color gamut},\ }\href {https://doi.org/10.1021/nl5048779} {\bibfield
  {journal} {\bibinfo  {journal} {Nano Letters}\ }\textbf {\bibinfo {volume}
  {15}},\ \bibinfo {pages} {3692} (\bibinfo {year} {2015})}\BibitemShut
  {NoStop}%
\bibitem [{\citenamefont {Becker}\ \emph {et~al.}(2018)\citenamefont {Becker},
  \citenamefont {Vaxenburg}, \citenamefont {Nedelcu}, \citenamefont {Sercel},
  \citenamefont {Shabaev}, \citenamefont {Mehl}, \citenamefont {Michopoulos},
  \citenamefont {Lambrakos}, \citenamefont {Bernstein}, \citenamefont {Lyons},
  \citenamefont {St\"{o}ferle}, \citenamefont {Mahrt}, \citenamefont
  {Kovalenko}, \citenamefont {Norris}, \citenamefont {Rainò},\ and\
  \citenamefont {Efros}}]{Becker2018}%
  \BibitemOpen
  \bibfield  {author} {\bibinfo {author} {\bibfnamefont {M.~A.}\ \bibnamefont
  {Becker}}, \bibinfo {author} {\bibfnamefont {R.}~\bibnamefont {Vaxenburg}},
  \bibinfo {author} {\bibfnamefont {G.}~\bibnamefont {Nedelcu}}, \bibinfo
  {author} {\bibfnamefont {P.~C.}\ \bibnamefont {Sercel}}, \bibinfo {author}
  {\bibfnamefont {A.}~\bibnamefont {Shabaev}}, \bibinfo {author} {\bibfnamefont
  {M.~J.}\ \bibnamefont {Mehl}}, \bibinfo {author} {\bibfnamefont {J.~G.}\
  \bibnamefont {Michopoulos}}, \bibinfo {author} {\bibfnamefont {S.~G.}\
  \bibnamefont {Lambrakos}}, \bibinfo {author} {\bibfnamefont {N.}~\bibnamefont
  {Bernstein}}, \bibinfo {author} {\bibfnamefont {J.~L.}\ \bibnamefont
  {Lyons}}, \bibinfo {author} {\bibfnamefont {T.}~\bibnamefont {St\"{o}ferle}},
  \bibinfo {author} {\bibfnamefont {R.~F.}\ \bibnamefont {Mahrt}}, \bibinfo
  {author} {\bibfnamefont {M.~V.}\ \bibnamefont {Kovalenko}}, \bibinfo {author}
  {\bibfnamefont {D.~J.}\ \bibnamefont {Norris}}, \bibinfo {author}
  {\bibfnamefont {G.}~\bibnamefont {Rainò}},\ and\ \bibinfo {author}
  {\bibfnamefont {A.~L.}\ \bibnamefont {Efros}},\ }\bibfield  {title} {\bibinfo
  {title} {Bright triplet excitons in caesium lead halide perovskites},\ }\href
  {https://doi.org/10.1038/nature25147} {\bibfield  {journal} {\bibinfo
  {journal} {Nature}\ }\textbf {\bibinfo {volume} {553}},\ \bibinfo {pages}
  {189–193} (\bibinfo {year} {2018})}\BibitemShut {NoStop}%
\bibitem [{\citenamefont {Yakunin}\ \emph {et~al.}(2015)\citenamefont
  {Yakunin}, \citenamefont {Protesescu}, \citenamefont {Krieg}, \citenamefont
  {Bodnarchuk}, \citenamefont {Nedelcu}, \citenamefont {Humer}, \citenamefont
  {De~Luca}, \citenamefont {Fiebig}, \citenamefont {Heiss},\ and\ \citenamefont
  {Kovalenko}}]{Yakunin2015}%
  \BibitemOpen
  \bibfield  {author} {\bibinfo {author} {\bibfnamefont {S.}~\bibnamefont
  {Yakunin}}, \bibinfo {author} {\bibfnamefont {L.}~\bibnamefont {Protesescu}},
  \bibinfo {author} {\bibfnamefont {F.}~\bibnamefont {Krieg}}, \bibinfo
  {author} {\bibfnamefont {M.~I.}\ \bibnamefont {Bodnarchuk}}, \bibinfo
  {author} {\bibfnamefont {G.}~\bibnamefont {Nedelcu}}, \bibinfo {author}
  {\bibfnamefont {M.}~\bibnamefont {Humer}}, \bibinfo {author} {\bibfnamefont
  {G.}~\bibnamefont {De~Luca}}, \bibinfo {author} {\bibfnamefont
  {M.}~\bibnamefont {Fiebig}}, \bibinfo {author} {\bibfnamefont
  {W.}~\bibnamefont {Heiss}},\ and\ \bibinfo {author} {\bibfnamefont {M.~V.}\
  \bibnamefont {Kovalenko}},\ }\bibfield  {title} {\bibinfo {title}
  {Low-threshold amplified spontaneous emission and lasing from colloidal
  nanocrystals of caesium lead halide perovskites},\ }\href
  {https://doi.org/10.1038/ncomms9056} {\bibfield  {journal} {\bibinfo
  {journal} {Nature Communications}\ }\textbf {\bibinfo {volume} {6}},\
  \bibinfo {pages} {8056} (\bibinfo {year} {2015})}\BibitemShut {NoStop}%
\bibitem [{\citenamefont {Utzat}\ \emph {et~al.}(2019)\citenamefont {Utzat},
  \citenamefont {Sun}, \citenamefont {Kaplan}, \citenamefont {Krieg},
  \citenamefont {Ginterseder}, \citenamefont {Spokoyny}, \citenamefont {Klein},
  \citenamefont {Shulenberger}, \citenamefont {Perkinson}, \citenamefont
  {Kovalenko},\ and\ \citenamefont {Bawendi}}]{Utzat2019}%
  \BibitemOpen
  \bibfield  {author} {\bibinfo {author} {\bibfnamefont {H.}~\bibnamefont
  {Utzat}}, \bibinfo {author} {\bibfnamefont {W.}~\bibnamefont {Sun}}, \bibinfo
  {author} {\bibfnamefont {A.~E.~K.}\ \bibnamefont {Kaplan}}, \bibinfo {author}
  {\bibfnamefont {F.}~\bibnamefont {Krieg}}, \bibinfo {author} {\bibfnamefont
  {M.}~\bibnamefont {Ginterseder}}, \bibinfo {author} {\bibfnamefont
  {B.}~\bibnamefont {Spokoyny}}, \bibinfo {author} {\bibfnamefont {N.~D.}\
  \bibnamefont {Klein}}, \bibinfo {author} {\bibfnamefont {K.~E.}\ \bibnamefont
  {Shulenberger}}, \bibinfo {author} {\bibfnamefont {C.~F.}\ \bibnamefont
  {Perkinson}}, \bibinfo {author} {\bibfnamefont {M.~V.}\ \bibnamefont
  {Kovalenko}},\ and\ \bibinfo {author} {\bibfnamefont {M.~G.}\ \bibnamefont
  {Bawendi}},\ }\bibfield  {title} {\bibinfo {title} {Coherent single-photon
  emission from colloidal lead halide perovskite quantum dots},\ }\href
  {https://doi.org/10.1126/science.aau7392} {\bibfield  {journal} {\bibinfo
  {journal} {Science}\ }\textbf {\bibinfo {volume} {363}},\ \bibinfo {pages}
  {1068} (\bibinfo {year} {2019})},\ \Eprint
  {https://arxiv.org/abs/https://www.science.org/doi/pdf/10.1126/science.aau7392}
  {https://www.science.org/doi/pdf/10.1126/science.aau7392} \BibitemShut
  {NoStop}%
\bibitem [{\citenamefont {Rain{\`o}}\ \emph {et~al.}(2022)\citenamefont
  {Rain{\`o}}, \citenamefont {Yazdani}, \citenamefont {Boehme}, \citenamefont
  {Kober-Czerny}, \citenamefont {Zhu}, \citenamefont {Krieg}, \citenamefont
  {Rossell}, \citenamefont {Erni}, \citenamefont {Wood}, \citenamefont
  {Infante},\ and\ \citenamefont {Kovalenko}}]{Raino2022}%
  \BibitemOpen
  \bibfield  {author} {\bibinfo {author} {\bibfnamefont {G.}~\bibnamefont
  {Rain{\`o}}}, \bibinfo {author} {\bibfnamefont {N.}~\bibnamefont {Yazdani}},
  \bibinfo {author} {\bibfnamefont {S.~C.}\ \bibnamefont {Boehme}}, \bibinfo
  {author} {\bibfnamefont {M.}~\bibnamefont {Kober-Czerny}}, \bibinfo {author}
  {\bibfnamefont {C.}~\bibnamefont {Zhu}}, \bibinfo {author} {\bibfnamefont
  {F.}~\bibnamefont {Krieg}}, \bibinfo {author} {\bibfnamefont {M.~D.}\
  \bibnamefont {Rossell}}, \bibinfo {author} {\bibfnamefont {R.}~\bibnamefont
  {Erni}}, \bibinfo {author} {\bibfnamefont {V.}~\bibnamefont {Wood}}, \bibinfo
  {author} {\bibfnamefont {I.}~\bibnamefont {Infante}},\ and\ \bibinfo {author}
  {\bibfnamefont {M.~V.}\ \bibnamefont {Kovalenko}},\ }\bibfield  {title}
  {\bibinfo {title} {Ultra-narrow room-temperature emission from single cspbbr3
  perovskite quantum dots},\ }\href
  {https://doi.org/10.1038/s41467-022-30016-0} {\bibfield  {journal} {\bibinfo
  {journal} {Nature Communications}\ }\textbf {\bibinfo {volume} {13}},\
  \bibinfo {pages} {2587} (\bibinfo {year} {2022})}\BibitemShut {NoStop}%
\bibitem [{\citenamefont {{M{\'e}ndez-Galv{\'a}n}}\ \emph
  {et~al.}(2025)\citenamefont {{M{\'e}ndez-Galv{\'a}n}}, \citenamefont
  {{Saucedo-Ch{\'a}vez}}, \citenamefont {{Medrano}}, \citenamefont
  {{Zuarez-Chamba}}, \citenamefont {{Garc{\'\i}a-Jomaso}}, \citenamefont
  {{Ord{\'o}{\~n}ez-Romero}}, \citenamefont {{Pirruccio}}, \citenamefont
  {{Camacho-Guardian}}, \citenamefont {{Soler-Illia}},\ and\ \citenamefont
  {{Lara-Garc{\'\i}a}}}]{Mendez2025}%
  \BibitemOpen
  \bibfield  {author} {\bibinfo {author} {\bibfnamefont {M.}~\bibnamefont
  {{M{\'e}ndez-Galv{\'a}n}}}, \bibinfo {author} {\bibfnamefont
  {Z.}~\bibnamefont {{Saucedo-Ch{\'a}vez}}}, \bibinfo {author} {\bibfnamefont
  {D.}~\bibnamefont {{Medrano}}}, \bibinfo {author} {\bibfnamefont
  {M.}~\bibnamefont {{Zuarez-Chamba}}}, \bibinfo {author} {\bibfnamefont
  {Y.~A.}\ \bibnamefont {{Garc{\'\i}a-Jomaso}}}, \bibinfo {author}
  {\bibfnamefont {C.~L.}\ \bibnamefont {{Ord{\'o}{\~n}ez-Romero}}}, \bibinfo
  {author} {\bibfnamefont {G.}~\bibnamefont {{Pirruccio}}}, \bibinfo {author}
  {\bibfnamefont {A.}~\bibnamefont {{Camacho-Guardian}}}, \bibinfo {author}
  {\bibfnamefont {G.~J.~A.~A.}\ \bibnamefont {{Soler-Illia}}},\ and\ \bibinfo
  {author} {\bibfnamefont {H.~A.}\ \bibnamefont {{Lara-Garc{\'\i}a}}},\
  }\bibfield  {title} {\bibinfo {title} {{Scalable Dip-Coated Bragg Mirrors for
  Strong Light-Matter Coupling with 2D Perovskites}},\ }\href
  {https://doi.org/10.48550/arXiv.2506.21726} {\bibfield  {journal} {\bibinfo
  {journal} {arXiv e-prints}\ ,\ \bibinfo {eid} {arXiv:2506.21726}} (\bibinfo
  {year} {2025})},\ \Eprint {https://arxiv.org/abs/2506.21726}
  {arXiv:2506.21726 [cond-mat.mes-hall]} \BibitemShut {NoStop}%
\bibitem [{\citenamefont {Georgakilas}\ \emph {et~al.}(2025)\citenamefont
  {Georgakilas}, \citenamefont {Tiede}, \citenamefont {Urbonas}, \citenamefont
  {Mirek}, \citenamefont {Bujalance}, \citenamefont {Cali{\`o}}, \citenamefont
  {Oddi}, \citenamefont {Tao}, \citenamefont {Dirin}, \citenamefont
  {Rain{\`o}}, \citenamefont {Boehme}, \citenamefont {Galisteo-L{\'o}pez},
  \citenamefont {Mahrt}, \citenamefont {Kovalenko}, \citenamefont {Miguez},\
  and\ \citenamefont {St{\"o}ferle}}]{Georgakilas2025}%
  \BibitemOpen
  \bibfield  {author} {\bibinfo {author} {\bibfnamefont {I.}~\bibnamefont
  {Georgakilas}}, \bibinfo {author} {\bibfnamefont {D.}~\bibnamefont {Tiede}},
  \bibinfo {author} {\bibfnamefont {D.}~\bibnamefont {Urbonas}}, \bibinfo
  {author} {\bibfnamefont {R.}~\bibnamefont {Mirek}}, \bibinfo {author}
  {\bibfnamefont {C.}~\bibnamefont {Bujalance}}, \bibinfo {author}
  {\bibfnamefont {L.}~\bibnamefont {Cali{\`o}}}, \bibinfo {author}
  {\bibfnamefont {V.}~\bibnamefont {Oddi}}, \bibinfo {author} {\bibfnamefont
  {R.}~\bibnamefont {Tao}}, \bibinfo {author} {\bibfnamefont {D.~N.}\
  \bibnamefont {Dirin}}, \bibinfo {author} {\bibfnamefont {G.}~\bibnamefont
  {Rain{\`o}}}, \bibinfo {author} {\bibfnamefont {S.~C.}\ \bibnamefont
  {Boehme}}, \bibinfo {author} {\bibfnamefont {J.~F.}\ \bibnamefont
  {Galisteo-L{\'o}pez}}, \bibinfo {author} {\bibfnamefont {R.~F.}\ \bibnamefont
  {Mahrt}}, \bibinfo {author} {\bibfnamefont {M.~V.}\ \bibnamefont
  {Kovalenko}}, \bibinfo {author} {\bibfnamefont {H.}~\bibnamefont {Miguez}},\
  and\ \bibinfo {author} {\bibfnamefont {T.}~\bibnamefont {St{\"o}ferle}},\
  }\bibfield  {title} {\bibinfo {title} {Room-temperature cavity
  exciton-polariton condensation in perovskite quantum dots},\ }\href
  {https://doi.org/10.1038/s41467-025-60553-3} {\bibfield  {journal} {\bibinfo
  {journal} {Nature Communications}\ }\textbf {\bibinfo {volume} {16}},\
  \bibinfo {pages} {5228} (\bibinfo {year} {2025})}\BibitemShut {NoStop}%
\bibitem [{\citenamefont {Polimeno}\ \emph {et~al.}()\citenamefont {Polimeno},
  \citenamefont {Coriolano}, \citenamefont {Mastria}, \citenamefont {Todisco},
  \citenamefont {De~Giorgi}, \citenamefont {Fieramosca}, \citenamefont
  {Pugliese}, \citenamefont {Prontera}, \citenamefont {Rizzo}, \citenamefont
  {De~Marco}, \citenamefont {Ballarini}, \citenamefont {Gigli},\ and\
  \citenamefont {Sanvitto}}]{Polimeno2024}%
  \BibitemOpen
  \bibfield  {author} {\bibinfo {author} {\bibfnamefont {L.}~\bibnamefont
  {Polimeno}}, \bibinfo {author} {\bibfnamefont {A.}~\bibnamefont {Coriolano}},
  \bibinfo {author} {\bibfnamefont {R.}~\bibnamefont {Mastria}}, \bibinfo
  {author} {\bibfnamefont {F.}~\bibnamefont {Todisco}}, \bibinfo {author}
  {\bibfnamefont {M.}~\bibnamefont {De~Giorgi}}, \bibinfo {author}
  {\bibfnamefont {A.}~\bibnamefont {Fieramosca}}, \bibinfo {author}
  {\bibfnamefont {M.}~\bibnamefont {Pugliese}}, \bibinfo {author}
  {\bibfnamefont {C.~T.}\ \bibnamefont {Prontera}}, \bibinfo {author}
  {\bibfnamefont {A.}~\bibnamefont {Rizzo}}, \bibinfo {author} {\bibfnamefont
  {L.}~\bibnamefont {De~Marco}}, \bibinfo {author} {\bibfnamefont
  {D.}~\bibnamefont {Ballarini}}, \bibinfo {author} {\bibfnamefont
  {G.}~\bibnamefont {Gigli}},\ and\ \bibinfo {author} {\bibfnamefont
  {D.}~\bibnamefont {Sanvitto}},\ }\bibfield  {title} {\bibinfo {title} {Room
  temperature polariton condensation from whispering gallery modes in cspbbr3
  microplatelets},\ }\href@noop {} {\bibfield  {journal} {\bibinfo  {journal}
  {Advanced Materials}\ }\textbf {\bibinfo {volume} {36}},\ \bibinfo {pages}
  {2312131}}\BibitemShut {NoStop}%
\bibitem [{\citenamefont {Porras}\ and\ \citenamefont
  {Tejedor}(2003)}]{porras}%
  \BibitemOpen
  \bibfield  {author} {\bibinfo {author} {\bibfnamefont {D.}~\bibnamefont
  {Porras}}\ and\ \bibinfo {author} {\bibfnamefont {C.}~\bibnamefont
  {Tejedor}},\ }\bibfield  {title} {\bibinfo {title} {Linewidth of a polariton
  laser: Theoretical analysis of self-interaction effects},\ }\href
  {https://doi.org/10.1103/PhysRevB.67.161310} {\bibfield  {journal} {\bibinfo
  {journal} {Phys. Rev. B}\ }\textbf {\bibinfo {volume} {67}},\ \bibinfo
  {pages} {161310} (\bibinfo {year} {2003})}\BibitemShut {NoStop}%
\end{thebibliography}%

\newpage

\begin{widetext}


\begin{center}
    {\LARGE \textbf{Supplementary Information}} \\[0.5cm]
    {\large {\bf Macroscopic coherence and vorticity in room-temperature polariton condensate confined in a self-assembled perovskite microcavity}} \\[0.3cm]
    {\normalsize Martin Montagnac$^{1,\dagger}$, Yesenia A. García Jomaso$^{1,\dagger}$, Emiliano Robledo Ibarra$^1$, Rodrigo Sánchez-Martínez$^1$, Moroni Santiago García, César L. Ordóñez-Romero$^1$, Hugo A. Lara-García$^{1,*}$, Arturo Camacho-Guardian$^{1,*}$, Giuseppe Pirruccio$^{1,*}$}
\end{center}

\begin{center}
    {
    $^{1}$ Instituto de Física, Universidad Nacional Autónoma de México, \\
    Apartado Postal 20-364, Ciudad de México C.P. 01000, México \\[0.3cm]
    $^{2}$ Instituto Nacional de Astrof\'isica, \'Optica y Electr\'onica, \\
    Calle Luis Enrique Erro 1, Sta. Ma. Tonantzintla, Puebla CP 72840, M\'exico
    }
\end{center}

 \thanks{$^{\dagger}$These authors contributed equally to this work.}
\date{\today}
\maketitle
\maketitle

In this Supplementary Information, we present the optical linear characterization of the sample.
For these measurements, we used samples obtained under slightly different fabrication parameters, briefly described below.

We also report complementary results obtained from other microcrystals belonging to the same sample as that discussed in the main text.
These microplatelets are fabricated under the same conditions and for this reason they differ only in their exact shape and lateral dimensions with respect to that presented in the main text. 

We discuss details of the experimental set-up used to perform the optical measurements presented in the main text, report complementary sanity checks for the interferometric measurements, discuss the quality of the fit of the spatial coherence map, show vortices obtained with a modified version of the interferometer and present the theoretical model based on the Gross-Pitaevskii equation.

\section*{Optical characterisation}
The optical absorption spectra were measured using a MAcylab UV1800PC UV--visible spectrophotometer.
Photoluminescence (PL) spectra were collected at room temperature under excitation with a $473\,\text{nm}$ laser, and the emitted light was analysed using the optical microscope described in the next section.
Spectroscopic ellipsometry was performed with a J.A. Woollam alpha-SE ellipsometer to determine the refractive index of the sample averaged over a spot of approximately $2\,\text{mm}$.

The sample was synthesised using a higher precursor concentration by increasing the initial mass in milligrams with respect to the fabrication procedure used for the sample described in the main text.
This way, we achieve larger surface coverage that enable reliable measurement of its linear optical properties (absorbance, PL and ellipsometry measurements).
It is worth noting that, because of the large illumination spot and the partial surface coverage of the substrate (approximately 50 \% filling), the refractive index is likely underestimated representing an average between the high-index material (crystals) and the surrounding low-index medium (air).

\begin{figure}[h!]
    \centering
    \includegraphics[width=0.6\linewidth]{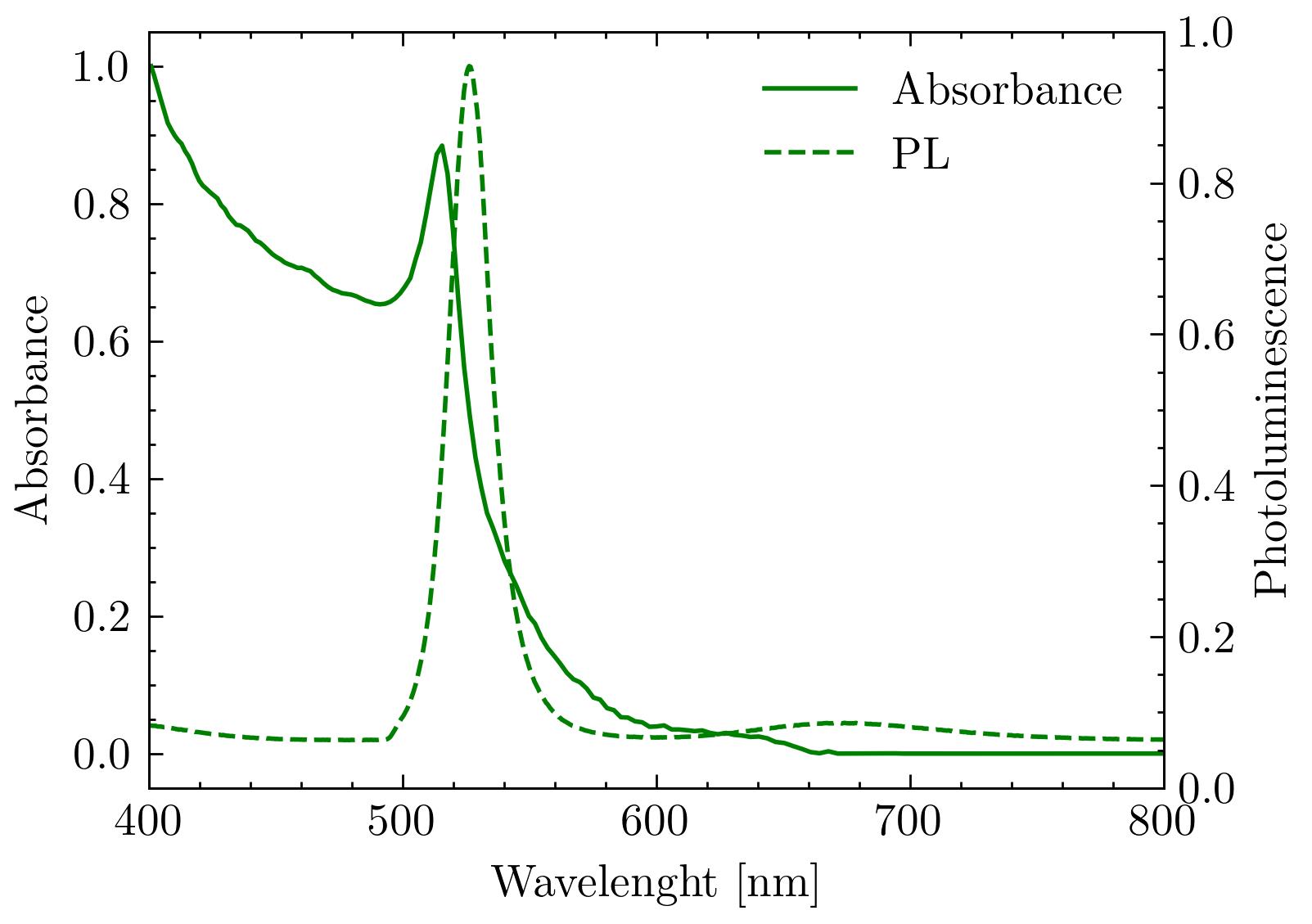}
    \caption{Normalised absorption and photoluminescence spectra of CsPbBr$_3$. The absorption spectrum exhibits a pronounced excitonic peak around 515 nm, while the PL emission is centred near 525 nm, indicating a Stokes shift of approximately 10 nm.  
    }
    \label{fig:abs}
\end{figure}
\begin{figure}[h!]
    \centering
    \includegraphics[width=0.6\linewidth]{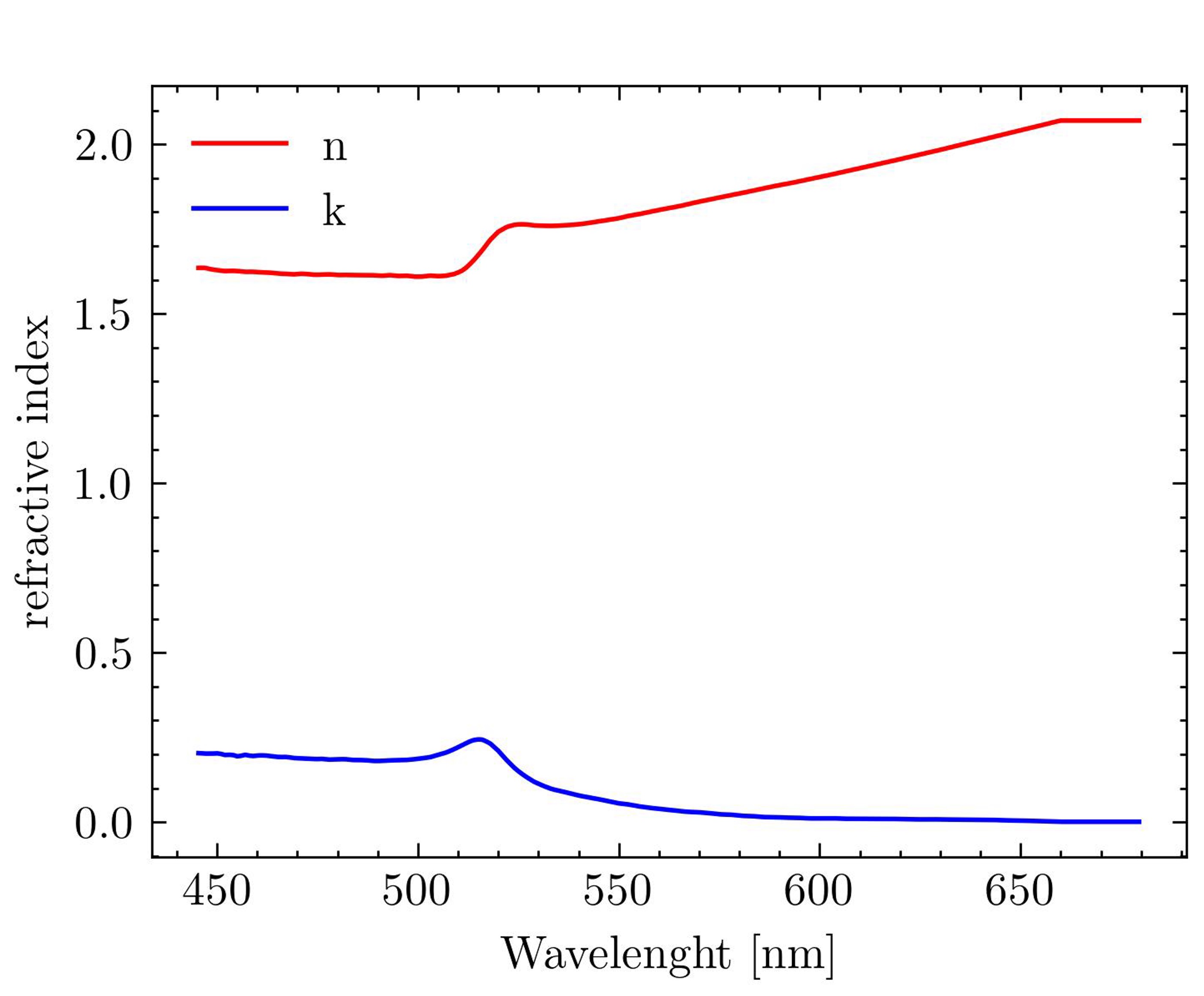}
    \caption{Complex refractive index as a function of wavelength, obtained by spectroscopic ellipsometry. The real part, $n$, indicated by the red curve, increases from approximately 1.6 at 450 nm to about 2.0 at 650 nm, showing normal dispersion. The peak around 520 nm in the imaginary part, $k$, shown with the blue curve, corresponds to excitonic absorption.
    }
    \label{fig:refrindex}
\end{figure}

\begin{figure}[h!]
    \centering
    \includegraphics[width=0.5\linewidth]{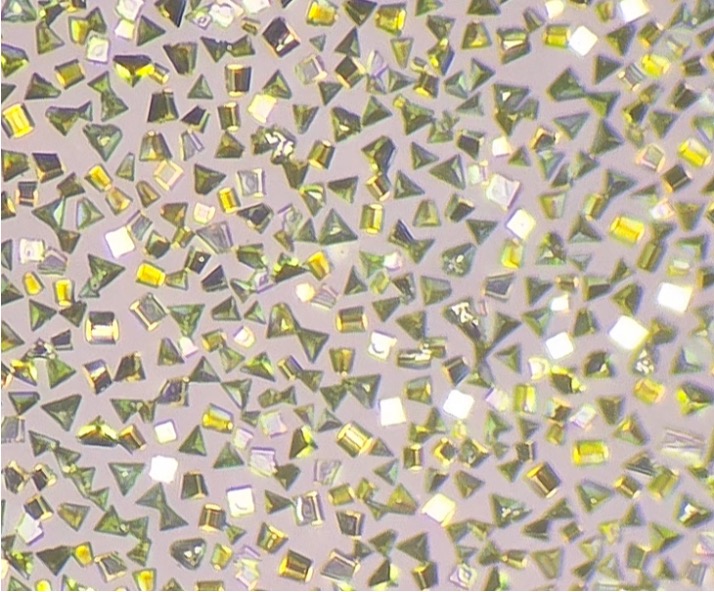}
    \caption{Optical microscopy image of the sample deposited on a SiO$_2$ substrate, showing a random distribution of triangular and square-shaped microcrystals. 
    }
    \label{fig:cristopt}
\end{figure}

\begin{figure}[h!]
    \centering
    \includegraphics{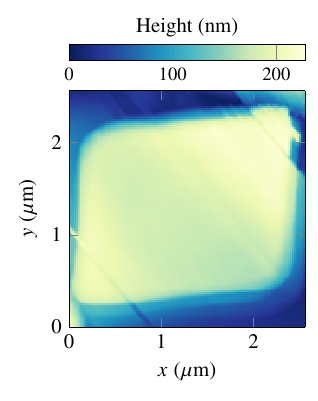}
    \caption{Atomic Force Microscopy image of a typical crystal showing an average height of $\approx200\,\text{nm}$.}
    \label{fig:cristopt}
\end{figure}

\section*{Spatial coherence measurement}

\subsection{Optical set-up}

The optical set-up used to measure the first-order spatial correlation function of single CsPbBr$_{3}$ microcrystals is shown in Figure. \ref{fig:setup}.
As discussed in the Methods section of the main text, the pulses emitted by the optical parametric amplifier are focused on the sample by means of a high numerical aperture microscope objective, which provides diffraction-limited resolution.
The beam diameter is expanded by means of two lenses in such a way that it overfills the objective entrance pupil.
The low repetition rate of the pump beam ensures complete relaxation of the system between each pulse.
The sample is mounted on an $xyz$ mechanical micropositioner.
The emitted light passes through a 4f-system incorporating an iris that permits spatial filtering.
High-pass filters are used to reject the pump light. Real-space imaging is obtained by focusing the filtered emitted light on the spectrometer by means of the tube lens, while k-space imaging is performed by adding a movable Bertrand lens before the tube lens.
Lasing is achieved by adjusting the focal spot size to the crystal area and varying the fluence by means of the software controlling the Pharos laser. 
To realize the spatial coherence measurements, the collimated emitted light is directed towards a Michelson interferometer comprising a 50:50 beam splitter, a motorized linear stage to achieve temporal overlap of the images traveling through the two arms, and a piezomirror to realize the phase scan. 
The hollow retroreflector ensured parallelism between the incident and reflected beam and provides a controllable lateral shift used to control the period of the interference fringes in the interferogram.
The recombined beam, formed by the two slightly non-collinear arm beams, is focused by the tube lens on the same sCMOS camera used for real-space imaging. 

\begin{figure}[h!]
    \centering
    \includegraphics[width=0.7\linewidth]{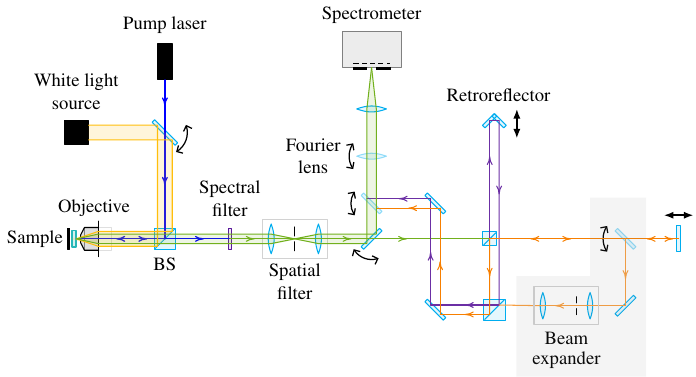}
    \caption{Sketch of the optical set-up used to measure the spectral and coherence properties of the single microplatelet.
    }
    \label{fig:setup}
\end{figure}

\begin{figure}[h!]
    \centering
    \includegraphics{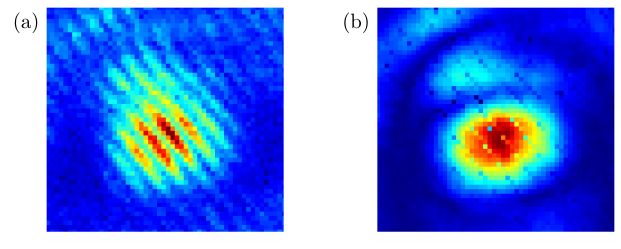}
    \caption{%
        (a) Pump beam interferogram showing a standard fringe pattern.
        (b) Spatial coherence map measured for the pump beam showing homogeneity over the region of high fringe visibility.
    }
    \label{fig:interf_g1_pump}
\end{figure}

To check the performance of the interferometer, the fitting algorithm and the phase extraction, we run the phase scan using the pump beam and replacing the sample with a commercial mirror.
Figure \ref{fig:interf_g1_pump}(a) displays the interferogram for a fixed OPD where high-contrast interference fringes as large as the FWHM of the focus point appear.
As expected, in this interferogram we do not observe the presence of fork-like dislocations.
The corresponding $g^{(1)}(\mathbf r,-\mathbf r)$ map is shown in Figure \ref{fig:interf_g1_pump}(b) where the high visibility of the fringes translates into a uniform, defect-free, large spatial coherence extending over the size of the focus.

\subsection{Vorticity from magnified arm image}

Optionally, a set of two lenses together with a $20\,\mu\text{m}$-pinhole can be placed in one of the two arms to select and magnify part of the image.
This allows overlap a region of the condensate of nearly constant spatial coherence with other parts of the condensate.
This region effectively works as a plane-wave probe unambiguously revealing the presence of vortices in the real-space image of the condensate and removing the possibility of double-counting.
In Figure~\ref{fig:bright_dark} we show a bright fork-like dislocation (a) resulting from the interference of the images propagating in the two arms of the interferometer (b and c).
The interference of the same magnified zone with a different region of the condensate is displayed in Figure~\ref{fig:bright_dark}(d-f) and reveals the presence of a dark dislocation. These two features are associated with vortices of opposite handedness.

\begin{figure}[h!]
    \centering
    \includegraphics{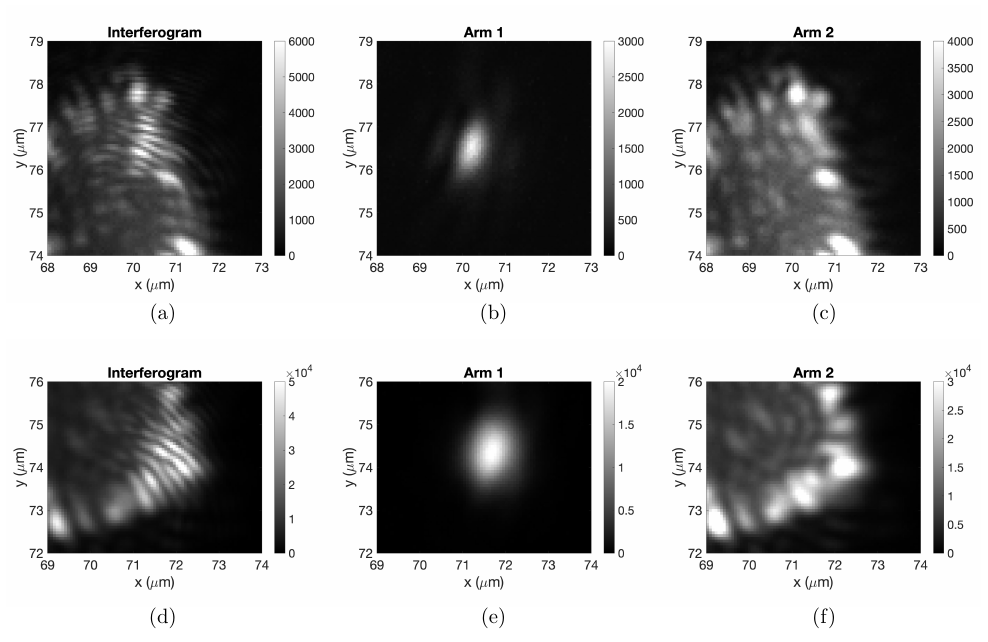}
    \caption{%
        (a) Interferogram showing a bright fork-like dislocation obtained by interfering a selected magnified area of the condensate of constant intensity.
        (b) with a non-magnified  PL image of the condensate.
        (c) Same as (a) but for an opposite handed vortex.
    }
    \label{fig:bright_dark}
\end{figure}

\section*{Fit quality of the spatial coherence map}

The quality of the fit performed on the intensity as a function of OPD for each pixel of the interferogram shown in Figure 4(a) of the main text is evaluated in two complementary ways:
(i) by computing the root mean square error (RMSE) for each pixel of the image and normalizing it to the mean amplitude of the corresponding oscillation;
(ii) by calculating the power contained in the main harmonic peak of the FFT spectrum for each data set, normalized to the total spectral power excluding the DC component. 
The frequency of this peak serves as the first guess for the least squares method.
The former quantity is plotted as a spatial map in Figure \ref{fig:rmse_norm_power_ratio}(b) where we see that the error of the fit throughout the entire cavity is around 15 $\%$.
The latter is plotted as a spatial map in Figure \ref{fig:rmse_norm_power_ratio}(a) and represents how close to a single cosine function each data set is.
We see that the experimental data are very well described by a single cosine function, which is a measure of the stability of the measurement.
This figure demonstrates the high-quality and consistency of the fit procedure, from which the $g^{(1)}(\mathbf r,-\mathbf r)$ map is obtained, over the majority of the crystal area.
In general, the regions where the accuracy of the fit is lower correspond to the regions of lower spatial coherence and the presence of vortices.

\begin{figure}[h!]
    \centering
    \includegraphics{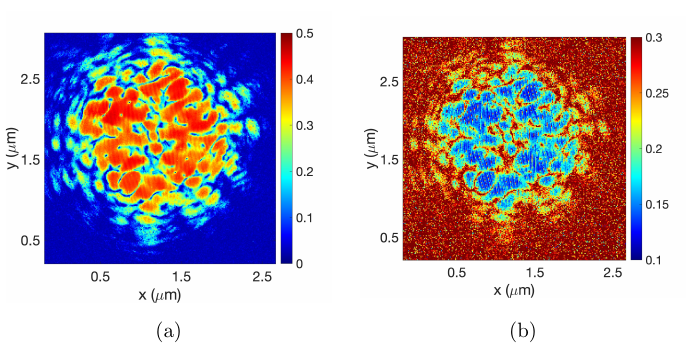}
    \caption{
    (a) Power ratio and (b) normalized RMSE for the fit performed in Figure 4 of the main text.
}
    \label{fig:rmse_norm_power_ratio}
\end{figure}

\subsection*{Fit of the period and phase }

 Besides the amplitude, from which the $g^{(1)}(\mathbf r,-\mathbf r)$ map is obtained, the least squares method outputs the spatial distribution of the initial phase and period of oscillations plotted in Figure \ref{fig:period_phi_fit}(b) and (a), respectively.
 The former further confirms the presence of fork-like dislocations, while the latter one provides a preliminary indication that the energy of the condensate is not constant over the crystal.
This is in agreement with the partial fragmentation of the condensate discussed in the main text and possibly constitutes a fast and easy alternative to spectral tomography.

\begin{figure}[h!]
\centering
\includegraphics{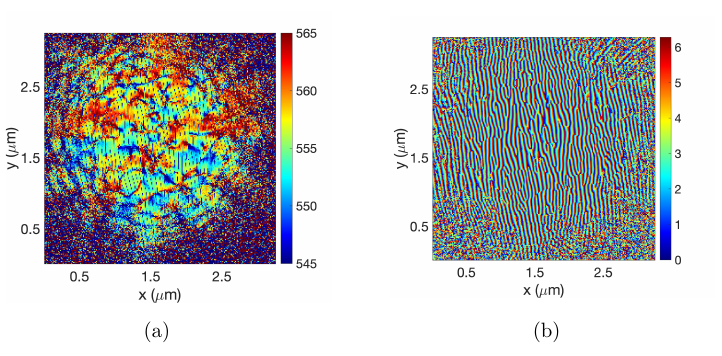}
\caption{%
    (a) Fitted period of the oscillations and (b) fitted phase offset complementing Figure 4 of the main text.
}
\label{fig:period_phi_fit}
\end{figure}

\section*{Temporal coherence}

The same optical set-up described in Figure \ref{fig:setup} can be used to measure the temporal coherence of the light emitted by a microcystal.
In this case, we delay one polariton laser pulse with respected to the other until no interference fringes are visible anymore in the interferogram.
This is accomplished by scanning the motorized retroreflector over a distance of several hundreds of micrometers and collecting the condensate spectrum from the whole crystal. Figure \ref{fig:temporal} displays the resulting gaussian-like curve from which the temporal coherence length is read-off as the FWHM.
It shows a measurement performed on a crystal possessing a spectrum very similar to the one discussed in the main text and for a power below saturation for which the spectrum can be reasonably approximated by a lorentzian curve.
In this case, the measured coherence length of 809 um, corresponding to $2.6\,\text{ps}$ and $0.3\,\text{nm}$, agrees well with the expected value obtained by taking the inverse of the FWHM of the spectrum.
However, for more complicated spectra exhibiting strong asymmetries or well-defined shoulders this measurement provides a clean estimation of the temporal coherence of the condensate.
By improving the stability of the motorized stage and minimizing vibrations, a larger signal-to-noise of the visibility curve measured for each pixel may be obtained.
This would allow reconstructing the spatial map of the temporal coherence, in the same spirit as the $g^{(1)}(\mathbf r,-\mathbf r)$ map.

\begin{figure}[h!]
   \centering
    \includegraphics[width=0.4\linewidth]{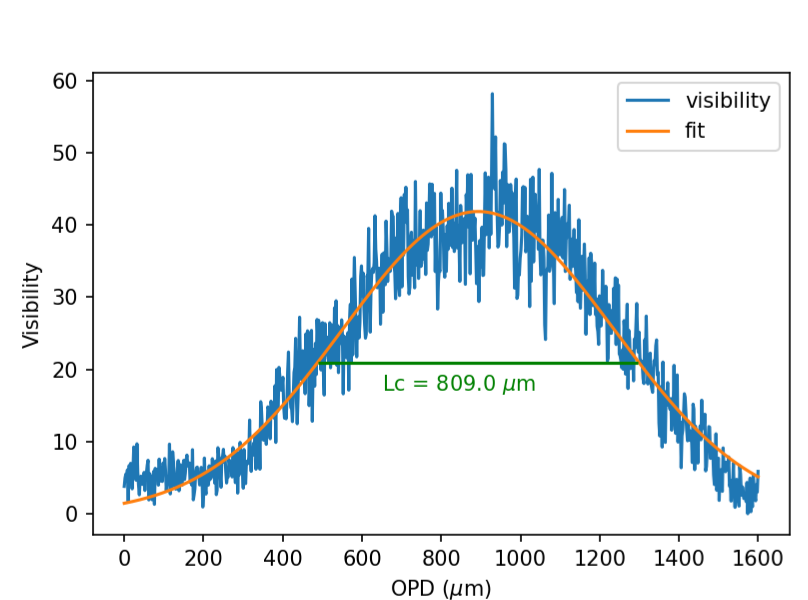}
    \caption{Temporal coherence measured interferometrically and gaussian fit from which the temporal coherence length is extracted. 
     }
    \label{fig:temporal}
 \end{figure}

\section*{Driven-dissipative Gross-Pitaevski equation}

To describe the dynamics of the polariton condensate in the presence of an incoherent reservoir and gain–loss processes, we employ a generalized open–dissipative Gross–Pitaevskii equation (GPE) coupled to a rate equation for the reservoir density $n_R(\mathbf{r}, t)$. The condensate order parameter $\psi(\mathbf{r}, t)$ evolves according to
\begin{align}
i\hbar \frac{\partial \psi(\mathbf{r}, t)}{\partial t} &= \left[ 
- \frac{\hbar^2 \nabla^2}{2m} 
+ g_c |\psi(\mathbf{r}, t)|^2 
+ V(\mathbf{r})
+\frac{i\hbar}{2}\left(R n_R(\mathbf{r}, t)-\gamma_c\right) 
 \right] \psi(\mathbf{r}, t),
\end{align}
\begin{equation}
\frac{\partial n_R(\mathbf{r}, t)}{\partial t} = P(\mathbf{r}) - \left( \gamma_R + R |\psi(\mathbf{r}, t)|^2 \right) n_R(\mathbf{r}, t).
\end{equation}

Here, $m$ is the effective mass of the lower polaritons, $g_c$ denotes the polariton–polariton interaction strength, and $V(\mathbf{r})$ is an external potential landscape. $\gamma_c$ and $\gamma_R$ are the decay rates of the condensate and reservoir, respectively, $P(\mathbf{r})$ is the spatial pump profile, and $R$ is the stimulated scattering rate from the reservoir into the condensate. The imaginary terms account for the driven–dissipative nature of the system: $-i \hbar \gamma_c/2$ represents radiative losses, and $i \hbar R n_R/2$ represents gain from stimulated scattering.

We analyse the steady-state properties within a single-mode approximation, replacing $\psi(\mathbf{r}, t) \rightarrow \psi(t)$ and $n_R(\mathbf{r}, t) \rightarrow n_R(t)$, and solve the resulting coupled ordinary differential equations. In steady state, the reservoir population is
\begin{equation}
     n_R = \frac{P}{\gamma_{R} + R |\psi|^2},
\end{equation}
and the condensate density is
\begin{equation}
|\psi|^2 =
\begin{cases}
\displaystyle \frac{P - P_{\mathrm{th}}}{\gamma_{c}}, & P > P_{\mathrm{th}} = \frac{\gamma_{c} \gamma_{R}}{R}, \\[1em]
0, & P < P_{\mathrm{th}} = \frac{\gamma_{c} \gamma_{R}}{R}.
\end{cases}
\end{equation}

\begin{figure}[h!]
    \centering
    \includegraphics[width=0.6\linewidth]{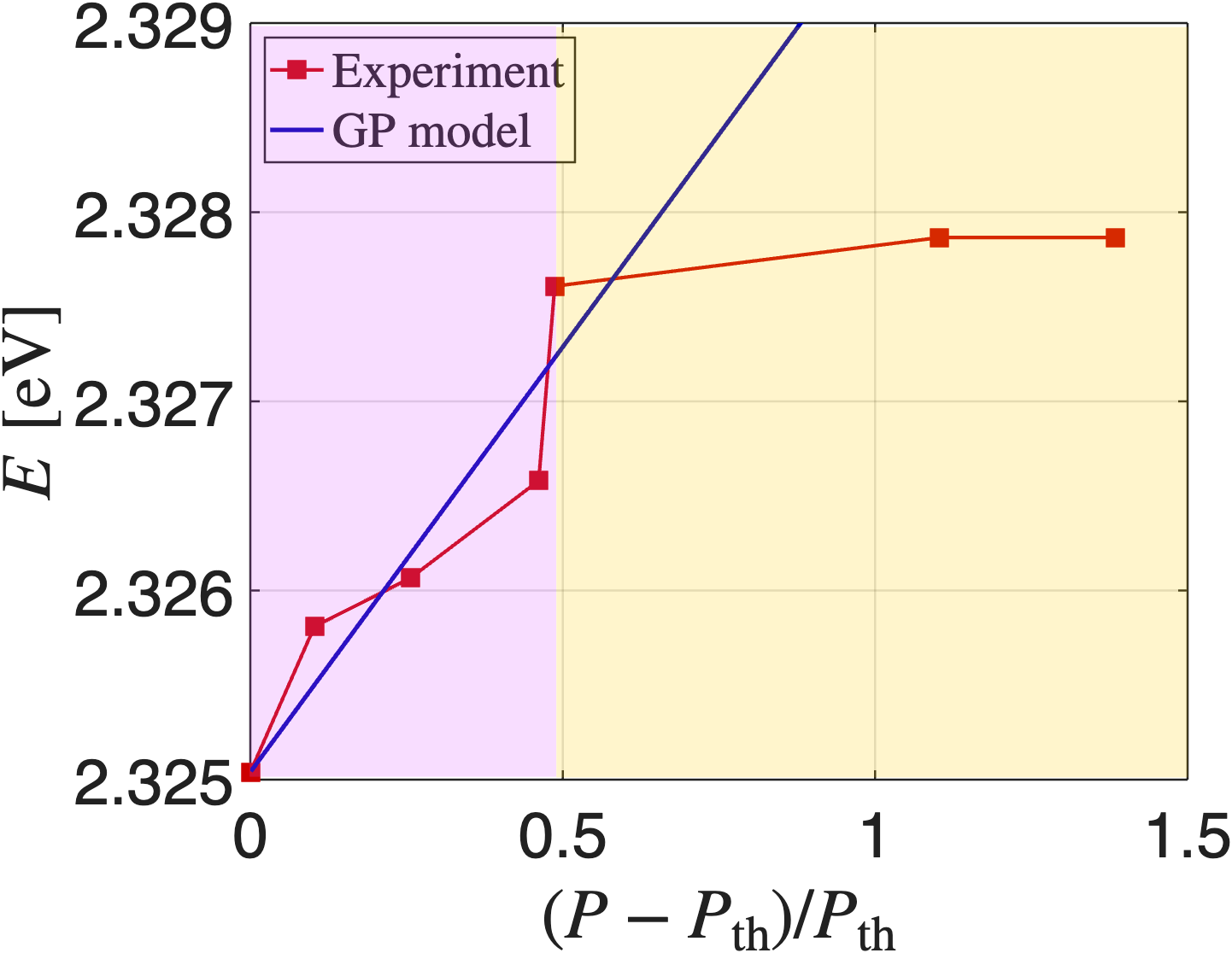}
    \caption{\textbf{Blueshift of the polariton condensate versus normalized pump power.} 
Red squares: experimentally extracted peak energies from photoluminescence spectra. 
Blue line: steady-state solution of the driven–dissipative Gross–Pitaevskii (GP) equation using independently measured parameters. 
In the low-excitation regime (purple), the blueshift grows linearly with pump intensity in quantitative agreement with the GP model, reflecting the linear scaling of condensate density above threshold. 
Beyond this regime (yellow), the experimental data saturate, signaling the breakdown of the simple GP description and the onset of additional nonlinear and dissipative processes.}
    \label{fig:TSM}
\end{figure}

Figure~\ref{fig:TSM} compares the measured condensate blueshift with the steady-state solution of the driven–dissipative GP equation. 
In the low-power regime just above threshold (purple shaded region), the experimental data follow a linear dependence on pump intensity, in excellent agreement with the GP model. 
This behaviour reflects the mean-field picture where the condensate density grows linearly with the excess pump power, and the blueshift is set by polariton–polariton interactions. 
At higher pump powers (yellow shaded region), the blueshift departs from linearity and saturates, indicating the breakdown of this simple GP description. 
This deviation marks the onset of nonlinear reservoir depletion, interaction renormalization, or other many-body effects beyond the scope of the mean-field model.
\end{widetext}
\end{document}